\documentclass[aps, prx, reprint, amsmath,amssymb,showpacs,floatfix,longbibliography, twocolumn, superscriptaddress,nofootinbib]{revtex4-2}
\usepackage{times} 
\usepackage{graphicx}
\usepackage{CJKutf8}
\usepackage{dcolumn}
\usepackage{bm}
\usepackage{color}
\usepackage{xcolor}
\usepackage{hyperref}
\usepackage{enumitem}
\usepackage{cancel}
\usepackage{amsfonts}
\usepackage{bbm}
\usepackage{xr-hyper}
\usepackage[caption=false]{subfig}
\usepackage{stmaryrd}
\usepackage{yhmath}
\usepackage{makecell}

\usepackage{comment}
\hypersetup{colorlinks=true,citecolor=blue,linkcolor=blue, urlcolor=blue}
\hypersetup{linktocpage}
\newcolumntype{M}[1]{>{\centering\arraybackslash}m{#1}}
\newcolumntype{N}{@{}m{0pt}@{}}
\usepackage{environ}

\bibpunct{[}{]}{,}{n}{}{}

\usepackage{amsfonts,amssymb,amsmath}
\usepackage[T1]{fontenc}
\usepackage{tikz}
\newcommand{\bra}[1]{\langle {#1} |}
\newcommand{\ket}[1]{ | {#1} \rangle}
\let\originalleft\left
\let\originalright\right
\renewcommand{\left}{\mathopen{}\mathclose\bgroup\originalleft}
\renewcommand{\right}{\aftergroup\egroup\originalright}

\newcounter{example}
\newenvironment{example}[1][]{\refstepcounter{example}\par\medskip
   \noindent \textbf{Example~\theexample. #1} \rmfamily}{\medskip}

\NewEnviron{eqs}{%
\begin{equation}\begin{split}
    \BODY
\end{split}\end{equation}
}

\begin{document}
\begin{CJK*}{UTF8}{gbsn}
\title{Tessellation codes: encoded quantum gates by geometric rotation}
\author{Yixu Wang}
\email{wangyixu@mail.tsinghua.edu.cn}
\affiliation{Institute for Advanced Study, Tsinghua University, Beijing, 100084, China}

\author{Yijia Xu (许逸葭)}
\email{yijia@umd.edu}
\affiliation{Joint Center for Quantum Information and Computer Science, University of Maryland, College Park,
Maryland 20742, USA}
\affiliation{Institute for Physical Science and Technology, University of Maryland, College Park, Maryland 20742, USA}

\author{Zi-Wen Liu}
\email{zwliu0@tsinghua.edu.cn}
\affiliation{Yau Mathematical Sciences Center， Tsinghua University, Beijing, 100084, China }

\date{\today}

\begin{abstract}
We utilize the symmetry groups of regular tessellations on two-dimensional surfaces of different constant curvatures, including spheres, Euclidean planes and hyperbolic planes, to encode a qubit or qudit into the physical degrees of freedom on these surfaces\textcolor{black}{, which we call tessellation codes.} 
We show that \textcolor{black}{tessellation} codes exhibit decent error correction properties by analysis via geometric considerations and the representation theory of the isometry groups on the corresponding surfaces.
Interestingly, we demonstrate how this formalism enables the implementation of certain logical operations through geometric rotations of surfaces in real space, \textcolor{black}{opening a new approach to logical quantum computation}.
We provide a variety of concrete constructions of such codes associated with different tessellations, which give rise to different logical groups. This formalism sheds a new light on quantum code and logical operation construction. 
\end{abstract}

\maketitle
\end{CJK*}

\section{Introduction}

In the pursuit of scalable quantum technologies such as quantum computing, quantum error correction is a challenging yet essential task.  Besides the robust storage of the encoded quantum information, efficient and robust manipulation of encoded quantum information through logical operations is a crucial but nontrivial problem.
Consequently,  designing quantum error-correcting (QEC) codes with desired logical gate sets emerges as a pressing problem that has garnered substantial attention, spanning a variety of important settings, including topological codes \cite{bombin2015gauge,kesselring2018boundaries,vasmer2019three,jochym2021four,zhu2022topological,zhu2023non,kesselring2024anyon}, product codes \cite{krishna2021fault,xu2024fast}, 
covariant codes  \cite{faist2020continuous,Woods_2020,zhou2021new,kubica2021using,liu2021quantum,kong2022near,PhysRevResearch.4.023107,liu2023approximate}, bosonic codes \cite{gkp,noh2020fault,baragiola2019all,calcluth2024sufficient,krastanov2015universal,heeres2017implementing,guillaud2019repetition,hillmann2020universal}, dynamical codes \cite{davydova2024quantum,kobayashi2024crosscap,fu2024errorcorrectiondynamicalcodes}, fusion-based quantum computing \cite{bartolucci2023fusion,bombin2023logical}.

In this work, we adopt a geometric perspective and consider a framework of quantum codes whose code states are extended in real space. 
\textcolor{black}{Spatial symmetries can then be harnessed as a resource.} The well-known Gottesman--Kitaev--Preskill (GKP) code~\cite{gkp} is underpinned by translation symmetry of flat space. We explore the rotational symmetries of surfaces with different curvatures and demonstrate how they can be utilized in QEC codes, in particular, enabling codes that support \textcolor{black}{implementations of logical quantum gates via geometric rotations.} While codes that allow rotation gates on spheres have been explored \cite{albert2020robust,gross2021designing,gross2023multispin,jain2024quantum}, we approach the code construction from the new perspective of regular triangle tessellations and propose a more general geometric formalism for constructing quantum codes, termed \emph{tessellation codes}, that encompasses surfaces with curvatures from positive to negative including spheres, Euclidean planes and hyperbolic \textcolor{black}{planes}. 
We note that some specific kinds of surface tessellation were used to construct microscopic models of holography~\cite{pastawski2015holographic,boyle2020conformal, boyle2024holographic} and quantum codes~\cite{li2023penrose} in different setups in previous studies.

Leveraging the connections between the group structures of the logical gate sets and the triangle groups of the tessellations, our tessellation codes enable representations of logical gate sets realizable by geometric rotations which are expected to be fault-tolerant and experimentally friendly, thus establishing a promising new formalism for logical quantum computing (see also Refs.~\cite{gross2021designing,gross2023multispin,denys2023,denys2023multimode, kubischta2023family,kubischta2023not,kubischta2024quantum} for different perspectives of such ``covariant encodings'' that have been under intensive study recently).   Relatedly, recent work has explored extending transversal gates through qubit lattice rotations within the framework of fold-transversal logical gates in a discrete-variable setting \cite{breuckmann2024fold}.
Specifically,  the codewords of tessellation codes are superpositions obeying patterns of surface tessellation and the logical gates correspond to rotational symmetries acting on the tessellation. \textcolor{black}{By choosing suitable tessellations, one can obtain codes with versatile logical gates (including single-qudit gates and entangling gates among code blocks) and error correction properties, yielding a systematic framework for geometric code and logical operation design.} 
To consolidate the general theory, we present a rich collection of examples produced by our formalism to illustrate the interplay among lattice symmetries, logical gate sets, and error correction.

Of particular interest are hyperbolic \textcolor{black}{planes}, whose exotic features may induce striking implications for coding theory and physics. This has been seen in classical codes~\cite{da2006signal}, the hyperbolic surface codes that overcome the Bravyi--Poulin--Terhal (BPT) bound \cite{bpt,breuckmann2016constructions,breuckmann2017hyperbolic}, band theory for non-abelian lattices \cite{maciejko2021hyperbolic,maciejko2022automorphic,lenggenhager2023non,boettcher2022crystallography}, and exotic phases of matter \cite{bienias2022circuit,stegmaier2022universality, urwyler2022hyperbolic,chen2024anderson, tummuru2024hyperbolic}. The present work provides the first examples of continuous-variable (CV) codes on hyperbolic \textcolor{black}{planes}. Especially, the rich (infinite) types of regular tessellations on hyperbolic \textcolor{black}{planes} enable us to realize a variety of logical gate sets, including the qudit Pauli group, the single-qubit Clifford group, the binary icosahedral group, and even the universal single-qubit gate set.  

\section{Group Structure}\label{sec:group_structure}
\textcolor{black}{In this work, we use group presentation $G=\langle A| R\rangle $ \cite{coxeter2013generators} which comprises two parts---generators $A$ and the relation of generators $R$---to describe a group $G$. For example, the qubit Pauli group has presentation}
\begin{equation}\label{eq:qubitPauli}
    \textcolor{black}{\mathcal{P}_{\text{qubit}}= }\langle X,Z |X^2=Z^2=(XZ)^4=\mathbbm{1}\rangle.
\end{equation}
For a qudit of general dimension $d\geq3$, \textcolor{black}{its Pauli group is generated by} $X=\sum_{j=0}^{d-1}\ket{j+1}\bra{j}$ and $Z=\sum_{j=0}^{d-1} \omega^j \ket{j}\bra{j}$, where $\omega=e^{i\frac{2\pi}{d}}$. The group presentation \textcolor{black}{of qudit Pauli group} is 
\begin{eqs}\label{eq:quditPauli}
   \mathcal{P}_{\text{qudit}}= \langle X,Z |X^d=Z^d=\Omega X \Omega^{-1}X^{-1}=\Omega Z \Omega^{-1}Z^{-1}=\mathbbm{1}\rangle.
\end{eqs}
Here $\Omega\equiv XZX^{-1}Z^{-1}$ is the \textcolor{black}{group} commutator of $X$ and $Z$. The last two relations \textcolor{black}{imply} that $\Omega$ is a central element. One can derive $\Omega^d=\mathbbm{1}$ with these relations. Therefore, $\Omega$ is identified as $e^{i\frac{2\pi}{d}}\mathbbm{1}$ in a unitary representation of the qudit Pauli group. \textcolor{black}{Furthermore, Eq.~\eqref{eq:quditPauli} leads to $(XZ)^d=\mathbbm{1}$ when $d$ is odd, and $(XZ)^{2d}=\mathbbm{1}$ when $d$ is even.}

There are various realizations of the single-qubit Clifford group that are equivalent up to global phases. An example presented in Ref.~\cite{denys2023multimode} is generated by
\begin{equation}\label{eq:Cliffordgates}
   S=\begin{pmatrix}
        e^{i\frac{\pi}{4}}&0\\0&e^{-i\frac{\pi}{4}}\\
    \end{pmatrix},~ U=\frac{1}{\sqrt{2}}\begin{pmatrix}
        e^{i\frac{\pi}{4}}&e^{i\frac{\pi}{4}}\\-e^{-i\frac{\pi}{4}}&e^{-i\frac{\pi}{4}}
    \end{pmatrix}.
\end{equation}
\textcolor{black}{These generators satisfy $S^{4}=U^{3}=(U S)^{2}=-\mathbbm{1}$, yielding a Clifford group of $48$ elements.} Its presentation is 
\begin{equation}\label{eq:quotient_singleClifford}
    \mathcal{C} =\langle S,U|~  S^8=U^6=(US)^{4}=S^4 U^3=S^4 (US)^2=\mathbbm{1}\rangle.
\end{equation}

\begin{figure}
    \centering    \includegraphics[height=0.23\textwidth]{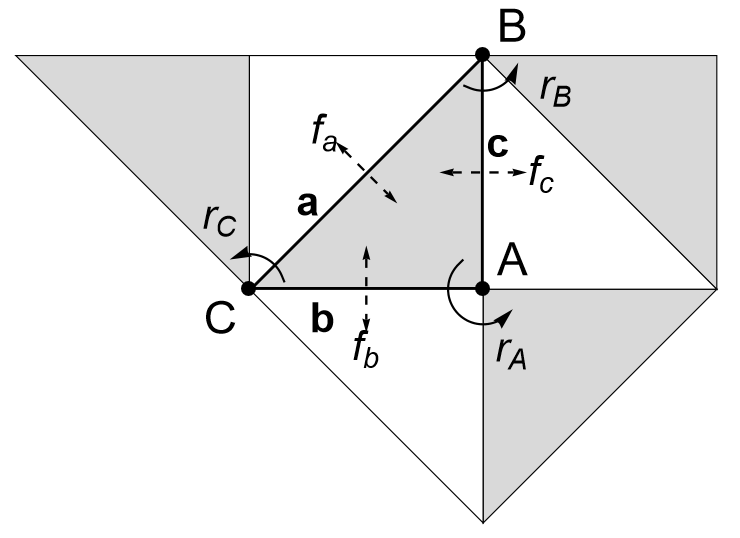}
    \caption{The $\{2,4,4\}$ tessellation on the plane is used to illustrate the action of the generators of the triangle and proper triangle groups. Generators of other $\{p,q,r\}$ groups act similarly geometrically. The dashed double-arrow lines represent reflections along the edges it crosses, and the solid single-arrow lines represent rotations. }
    \label{fig:TriGroup}
\end{figure}

Regular triangle tessellations of two-dimensional surfaces \textcolor{black}{ are characterized by three positive integers $\{p,q,r\}$, which represent three internal angles $\{\frac{2\pi}{p}, \frac{2\pi}{q}, \frac{2\pi}{r}\}$ of the unit triangle. These three integers also determine the symmetry group of the tesselation, known as the triangle group.}
A triangle group \cite{magnus1974noneuclidean} has three generators which are reflections along each \textcolor{black}{edge} of a triangle, \textcolor{black}{labeled by their corresponding edges as $f_a$, $f_b$, $f_c$.} \textcolor{black}{We denote the} triangle group \textcolor{black}{associated with} \textcolor{black}{the tessellation $\{p,q,r\}$ as} $\Delta (p,q,r)$ \textcolor{black}{with the presentation},
\begin{equation}\label{eq:triangle}
   \langle f_a,f_b,f_c|~f_a^2=f_b^2=f_c^2=(f_b f_c)^p=(f_c f_a)^q=(f_a f_b)^r=\mathbbm{1}\rangle.
\end{equation}
An index-$2$ subgroup is called the proper triangle group, denoted as $\Bar{\Delta}(p,q,r)$. \textcolor{black}{Its elements are products of even numbers of reflection generators}. By redefining $r_A\equiv f_b f_c$, $r_B\equiv f_c f_a$, $r_C\equiv f_a f_b=(r_A r_B)^{-1}$, the presentation \textcolor{black}{ of the proper triangle group} can be written as 
\begin{equation}\label{eq:vonDyck}
   \textcolor{black}{\bar{\Delta}(p,q,r)=}\langle r_A, r_B|~r_A^p=r_B^q=(r_A r_B)^r=\mathbbm{1}\rangle.
\end{equation}
Here $r_A= f_b f_c$ is the rotation around the angle $A$ defined by sides $b$ and $c$ in the direction from $c$ to $b$ by $\ \frac{2\pi}{p}$. $r_B$ and $r_C$ have similar geometric meanings, as depicted in Figure.~\ref{fig:TriGroup}.

The integers label the types of the tessellations. Their reciprocal sum is related to the curvature of the surfaces,
\begin{equation}\label{eq:pqrcurvature}
    \frac{1}{p}+\frac{1}{q}+\frac{1}{r}~~\begin{cases}
        >1&\implies \text{sphere},\\
        =1&\implies \text{Euclidean plane},\\
        <1&\implies \text{hyperbolic plane}.\\
    \end{cases}
\end{equation}

\textcolor{black}{We observe that single-qudit Pauli, Clifford, and proper triangle groups are all generated by two generators and specific relations that certain products of generators equal to the identity element.} The structural similarities between the presentation of proper triangle groups \eqref{eq:vonDyck} and the Pauli group \eqref{eq:quditPauli} or the single-qubit Clifford group \eqref{eq:quotient_singleClifford} motivate us to realize logical qubit gates via geometric manipulations.

The presentations of the Pauli or Clifford groups have extra relations compared to their \textcolor{black}{proper} triangle group counterparts. From an algebraic perspective, imposing extra relations $R$ is described as the quotient of the triangle group by the normal subgroup called the normal closure of $R$. The normal closure is generated by the generators of the form $R^{\bar\Delta}=\{g^{-1}rg,~g\in \bar\Delta,~r\in R\}$.
Geometrically, these extra relations \textcolor{black}{impose identifications among lattice points in the tessellation, thereby defining fundamental unit cells of codewords.} The normal closure is the translation group in lattice theory and is often denoted as $\Gamma$. In the context of quantum error correction theory, it plays a similar role to the stabilizer group in the non-commuting case, so we call $\Gamma$ the generalized stabilizer group. \textcolor{black}{The logical group $G$ is then related to the proper triangle group $\bar{\Delta}$ and the generalized stabilizer group $\Gamma$ by}
\begin{eqs} 
\label{eq:logicalgroup}
    &G=\bar{\Delta}/\Gamma.
\end{eqs}

\section{General Formalism}\label{sec:general_formalism}

\subsection{Encoding strategy}\label{sec:encoding}
We now explain the general framework for constructing \textcolor{black}{tessellation codes} that enable geometric realizations of logical gates. The physical degree of freedom is a free massless two-dimensional boson on the corresponding surfaces. The physical Hilbert space $\mathcal{H}_P$ corresponds to the space of wavefunctions defined on these surfaces. The logical degree of freedom is a qubit or a qudit, which can be understood as a virtual system that hosts the logical group actions as we shall discuss. The corresponding logical space $\mathcal{H}_L$ is the $d$-dimensional vector space for qudits. \textcolor{black}{Our goal is to construct a covariant encoding for group $G$, where a $d$-dimensional unitary representation $\rho_L(g)$ acting on the logical space $\mathcal{H}_L$ is implemented by a geometric rotation $\rho(g)$ acting on the physical space $\mathcal{H}_P$ for $\forall g\in G$.} For the Euclidean and hyperbolic cases, each logical $g$ can be written as $\rho(g)=\Gamma g_0=\sum_{\gamma\in \Gamma} \gamma g_0$, for an arbitrary coset representative $g_0$ under the quotient of Eq.~\eqref{eq:logicalgroup}. 

We now construct the positional configurations of the \textcolor{black}{codewords}. As the encoding map is covariant for $G$, it should involve the structure of a linear map intertwining between representations $V=\sum_{g\in G}\rho_L(g)^{-1}\otimes\rho(g)$.  The general encoding map $\mathcal{E}$ takes the form in Ref.~\cite{denys2023multimode}. 
The \textcolor{black}{codeword} $\ket{\Bar{k}} \in \mathcal{H}_P$, which represents the logical state $\ket{k}\in \mathcal{H}_L$ to be encoded, can be written as 
\begin{equation}\label{eq:encodingmap}
\ket{\Bar{k}}\equiv\mathcal{E}(\ket{k})\propto \sum_{g\in G}\bra{\Sigma} \rho_L(g)^\dagger\ket{k}\rho(g) \ket{\text{init}}.
\end{equation}
Here the states $\ket{\Sigma}\in \mathcal{H}_L$ and $\ket{\text{init}}\in \mathcal{H}_P$ are chosen such that $\bra{\Sigma}V\ket{\text{init}}\neq 0$.
\textcolor{black}{In most cases, we choose $\ket{\Sigma}=\ket{0}\in \mathcal{H}_L$.} For simplicity, we take the \textcolor{black}{initial state} $\ket{\text{init}}$ as a delta function state localized at a given point $p_i$ on the two-dimensional surface. We write $\ket{\text{init}}=\ket{p_i}$ hereafter \textcolor{black}{, which will be parameterized in spherical, Cartesian, and hyperbolic polar coordinates for sphere, Euclidean, and hyperbolic planes respectively.} The collection of delta functions for a given codeword is called its configuration. $\ket{p_i}$ can be chosen arbitrarily inside or on the edges of a unit triangle (see Fig. \ref{fig:TriGroup}) for code construction, except for the vertices. The collection of delta functions for a given codeword is called its configuration. Different choices of $\ket{p_i}$ induce different code constructions. Whether $\ket{p_i}$ can be put on vertices of the triangle and how choices of $\ket{p_i}$ affect the code performance are discussed in detail in Appendix.~\ref{app:constellation}.

\subsection{Error model and correction}\label{sec:error_model}
We discuss error models on each type of surface and introduce parameters to characterize the code performance. The position error is modelled by an orientation-preserving isometric transformation of the plane and reflects the imprecision of logical operation implementations or unexpected vibration of the system. We define the robustness against position error by resolution $d_x$, which is half of the minimum distance between two points in the code configurations. The momentum errors are modelled by the eigenfunctions of the Laplacian operators on each surface type, which are labelled by two indices for two-dimensional surfaces and form bases of the irreducible representations of the isometry group. Physically, the eigenvalues are the kinetic energies of the modes. Momentum errors model the unexpected energy excitations of the physical system.  Mathematically, they are bases of irreducible representations of the isometry group. In Table \ref{tab:errortype}, we list different types of position and momentum errors on corresponding surfaces. We denote an error operator as $\hat E_{r,n}$, in which $r$ labels the representation and $n$ labels the $n$-th basis of the vector space carrying the representation.
The error correction properties of codes from our framework can be analyzed based on these interpretations. In the main text, we highlight the logical operation aspects, \textcolor{black}{while a more detailed discussion on error correction is provided Appendix.~\ref{app:constellation} and \ref{app:momentum}}

The code's performance against position errors can be naturally characterized by geometric parameters. One such parameter is one-half of the minimal distance among pairs of adjacent points, which we call the resolution $d_x$. For an isometry $g$ which transforms a point up to a finite distance, $\|p_i -g p_i\|\leq d_g< \infty$, if $d_g < d_x$, then the error $g$ is correctable. The resolution $d_x$ indeed depends on codeword configurations, through the choice of $\ket{p_i}$ in Eq.~\eqref{eq:encodingmap}. In Appendix.~\ref{app:constellation}, we show how to determine the $|p_i\rangle$ that gives the optimal resolution configuration of a given tessellation.

Correcting position error is determining an unknown isometry. In principle, there are at most two fixed points for isometries on a sphere and at most one for isometries on an Euclidean plane or a hyperbolic plane (excluding points at infinity). Therefore, we can measure two points (or three points on a sphere if the first two measurement results are antipodal) and compare their positions before and after the isometry. The unknown isometry should be solved up to an element of the generalized stabilizer group. We can then apply the inverse isometry to correct the error. Every measurement gives one position and we need to at least measure twice.

\textcolor{black}{The} configurations of noiseless codewords are known \textcolor{black}{and} denoted as set $L$. The syndrome measurement gives the positions of two points $p_1^{(m)}$, $p_2^{(m)}$ of the noisy state. The error correction amounts to determining their original positions and the error $g_e$. We examine different ansatz of isometries that take the two positions back to one pair of points in the original configuration $L$. We denote all such isometries as the set  $Q=\{g|g p_1^{(m)},g p_2^{(m)}\in L\}$.  If a probability distribution $\mu$ of the noise group elements is given, we may use the maximal probability decoder to find the most probable error $g_e$: 
\begin{equation}
    g_e=\underset{g:  g\in Q }{\text{argmax}}~\mu(g).
\end{equation}

The momentum errors are basis of irreducible representations of the isometry group. We denote momentum error operator as $\hat E_{r,n}$, in which $r$ labels the representation and $n$ labels the $n$-th basis of the vector space carrying the representation. In correspondence with Table. \ref{tab:errortype}, $r\to \ell$ and $n\to m$ for the spherical errors, $r\to (k_x, k_y)$ for the Euclidean errors, and $r\to r$, $n\to n$ for the hyperbolic errors. 
The analysis of the Knill--Laflamme error correction conditions \cite{knill_laflamme} of momentum errors can be unified with the help of the representation theory of the isometry groups. Consider the pairs of errors $\hat{E}_{r_1,n_1}$, $\hat{E}_{r_2,n_2}$, if $(r_1,n_1)=(r_2,n_2)$, the error correction condition is always satisfied. For the cases where $(r_1,n_1)\neq(r_2,n_2)$, most of them also satisfy the error correction condition, exactly due to the consequence of the representation theory of the generalized stabilizer group by Schur's lemma. 
It is straightforward to calculate the violating pairs for the codes on the sphere and Euclidean plane but harder for those on the hyperbolic plane. We conjecture that the \textcolor{black}{error correction performance against} momentum error can be characterized by the first non-trivial eigenvalue of the Laplacian of the compact manifold after the quotient of the generalized stabilizer group. We collect the representation-theoretic analysis of error correction conditions of momentum errors in Appendix.~\ref{app:momentum}.

\section{Case studies}\label{sec:examples}
In the following, we introduce some representative examples of codes associated with spherical, planar, and hyperbolic tessellations derived from the above general framework, with certain groups of logical operations realized by geometric rotations. 
There are infinitely many different regular tessellations on the hyperbolic plane enabling us to realize a wide range of logical gate sets. We only present three examples as proof of principle. We describe each code by the type of tessellation, the generalized stabilizers, the logical operator generators, and the code performance against position and momentum errors. To lighten the notation, we take the curvature radius of both sphere and hyperbolic plane to be $1$. For the Euclidean plane, the side length of the isosceles right triangle is $1$. So is the side length of the equilateral triangle. We summarize the parameters and key information of all codes in Table \ref{tab:codedata}. In the main text, we present the codewords configurations of the examples with a particular choice of $\ket{p_i}$. The general codewords and their relations to other codes are given in Appendix \ref{app:code_construction}.

\begin{widetext}
\begin{center}
  \begin{table}[htpb]
    \centering
    \begin{tabular}{|c|c|c|c|}
    \hline
     & Sphere & Euclidean plane & Hyperbolic plane\\
    \hline
    Isometry group & SO(3) & E(2) & PSL(2,$\mathbb{R}$)\\
    \hline
    Position errors & Rotation along any axis & Translations, rotations & Hyperbolic translations, rotations\\
    \hline
    Momentum errors & $Y^m_\ell(\hat\theta,\hat\phi)=P_{\ell}^m(\cos \hat{\theta})e^{i m \hat\phi}$& $\exp{(i (k_x \hat{x}+ k_y \hat{y}))}$ & $P_{-\frac 12+i s}^n(\cosh \hat{\rho})e^{i n \hat\phi}$\\
    \hline
    \end{tabular}
    \caption{Different types of position and momentum errors in the table. Position errors are elements of orientation-preserving isometry groups. Momentum errors are the eigenfunctions of the Laplacians of the corresponding surfaces. They are also basis of unitary representations of the isometry group.}
    \label{tab:errortype}
\end{table}  
\begin{table}[htpb]
    \centering
    \begin{tabular}{|c|c|c|c|c|c|c|}
    \hline
    &Example \ref{example:224XZ}& Example \ref{example:244XZ}& Example \ref{example:333XZ}& Example \ref{example:555XZ}& Example
    \ref{example:468Clifford}& Example \ref{example:4352I}
    \\
    \hline
    Designated logical group &Qubit Pauli& Qubit Pauli & Qutrit Pauli &  $\mathbb{Z}_5$ qudit Pauli & Qubit Clifford & Binary icosahedral \\
    \hline
    Surface type& Sphere & Euclidean plane & Euclidean plane & Hyperbolic plane & Hyperbolic plane & Hyperbolic plane
    \\
    \hline
    Triangle tessellation & $\{2,2,4\}$& $\{2,4,4\}$& $\{3,3,3\}$& $\{5,5,5\}$&$\{6,4,8\}$&$\{4,3,5\}$\\
    \hline
    & $r_A\to Z$&$r_A\to X$&$r_A\to Z$&$r_A\to Z$&$r_A\to U $&$r_A \to \Phi F^{-1}$\\
    
    Generator identifications & $r_B\to X$ & $r_B\to Z$ & $r_B\to X$ & $r_B\to X$ &$r_B\to (US)^{-1}$&$ r_B\to -F$\\
    & $r_C\to XZ$&$r_C\to (XZ)^{-1}$&$r_C\to (ZX)^{-1}$&$r_C\to (ZX)^{-1}$&$r_C\to S$& $r_C\to -\Phi^{-1}$\\
    \hline
    Extra relations & $-$ & $r_B^2$ & \makecell{ $\Omega r_A\Omega^{-1} r_A^{-1}$,\\$\Omega r_B\Omega^{-1} r_B^{-1}$}&\makecell{ $\Omega r_A\Omega^{-1} r_A^{-1}$,\\$\Omega r_B\Omega^{-1} r_B^{-1}$} & $r_B^2 r_A^3$, $r_B^2 r_C^4$ & $r_B^2r_A r_B^2 r_A^{-1}$ \\
    \hline
    State configuration & Figure \ref{fig:224XZsphere} & Figure \ref{fig:244XZplane} &Figure \ref{fig:333XZplane} & Figure \ref{fig:555QuditXZ} & Figure \ref{fig:468Cliffordhyperbolic}& Figure \ref{fig:4352Ihyperbolic} \\
    \hline
    Resolution $d_x$ &$\frac{1}{2}\arccos{\frac{1}{3}}$ & $\frac 12$ & $\frac{\sqrt{3}}{2}$ & 1.6169 &0.6605& 0.5011\\  \hline
    \end{tabular}
    \caption{Different properties of the codes constructed in this work. We denote $\Omega\equiv r_B r_A r_C$, which is identified as $XZX^{-1}Z^{-1}$, the commutator of the Pauli $X$ and $Z$ operator. }
    \label{tab:codedata}
\end{table}
\end{center}
\end{widetext}

\begin{example}\label{example:224XZ}
First, we utilize the sphere's $\{2,2,4\}$ tessellation to realize a qubit code with logical Pauli operations realized by rotations. The $\bar\Delta(2, 2, 4)$ is isomorphic to the qubit Pauli group \eqref{eq:qubitPauli}. The most general form of the codewords is presented in Eq. \eqref{eq:224_codewords}. In Figure \ref{fig:224XZsphere}, the configuration corresponds to the specific choice $\theta_0=\arccos\frac{1}{\sqrt{3}}$, $\phi_0=\frac{\pi}{4}$. They are on the vertices of a cube. The logical operations are implemented as rotations along different axes. The logical $Z$ is rotating around the $x$ axis by $\pi$. The logical $X$ is rotating around the diagonal of the $x, y$ axis by $\pi$. The logical $XZ$ is to rotate around the $z$ axis counterclockwise by $\frac{\pi}{2}$.

The position error on the sphere is a rotation $R(\theta,\phi,\alpha)$ around the axis through the point $(\theta,\phi)$ by an angle $\alpha$. 
Therefore, for any two rotations $R_1(\theta_1,\phi_1,\alpha_1)$ and $R_2(\theta_2,\phi_2,\alpha_2)$ with $\alpha_1, \alpha_2 < \frac{1}{2}\arccos{\frac{1}{3}}$, the off-diagonal KL condition
$\bra{\bar 0} R_2^\dagger R_1 \ket{\bar 1}=\bra{\bar 1} R_2^\dagger R_1 \ket{\bar 0}=0$ is always satisfied. For the diagonal condition $\bra{\bar 0} R_2^\dagger R_1 \ket{\bar 0},~\bra{\bar 1} R_2^\dagger R_1 \ket{\bar 1}$, they are also $0$, unless the cases in which the axis of the composite rotation $R_2^\dagger R_1$ exactly passes through one of the configuration point in codeword $\ket{\bar 0}$ or $\ket{\bar 1}$. However, in these special cases, because the antipodal points on the rotational axis belong to $\ket{\bar 0}$ and $\ket{\bar 1}$ respectively for the configuration in Figure \ref{fig:224XZsphere}, the diagonal error correction is still satisfied. Therefore, this code can correct the set of rotations $R(\theta,\phi,\alpha)$ with arbitrary $(\theta,\phi)$ and $\alpha <  \frac{1}{2}\arccos{\frac{1}{3}}$.

The momentum errors are the spherical harmonics $Y^m_
\ell:=Y^m_\ell(\hat\theta,\hat\phi)=P_{\ell}^m(\cos \hat{\theta})e^{i m \hat\phi}$. Because the logical $\ket{\bar 0}$ and $\ket{\bar 1}$ have no spatial overlap, the off-diagonal error correction conditions are always satisfied. For the diagonal part, we evaluate $\bra{\bar 0}  Y^{m_1 \dagger}_{\ell_1}  Y^{m_2}_{\ell_2}\ket{\bar 0},~\bra{\bar 1}  Y^{m_1 \dagger}_{\ell_1}  Y^{m_2}_{\ell_2}\ket{\bar 1}$ in Appendix \ref{sec:224XZmomentum}. The KL condition is only violated when $\ell_1+\ell_2$ is odd and $(m_1+m_2)\mod 4=2$. The lowest uncorrectable pair of the errors is $Y_1^{0\dagger}Y_2^{\pm 2}$. Therefore, this code can correct all the momentum errors $Y_\ell^m$ with $\ell \leq 1$.

\begin{figure}
    \centering    \includegraphics[height=0.35\textwidth]{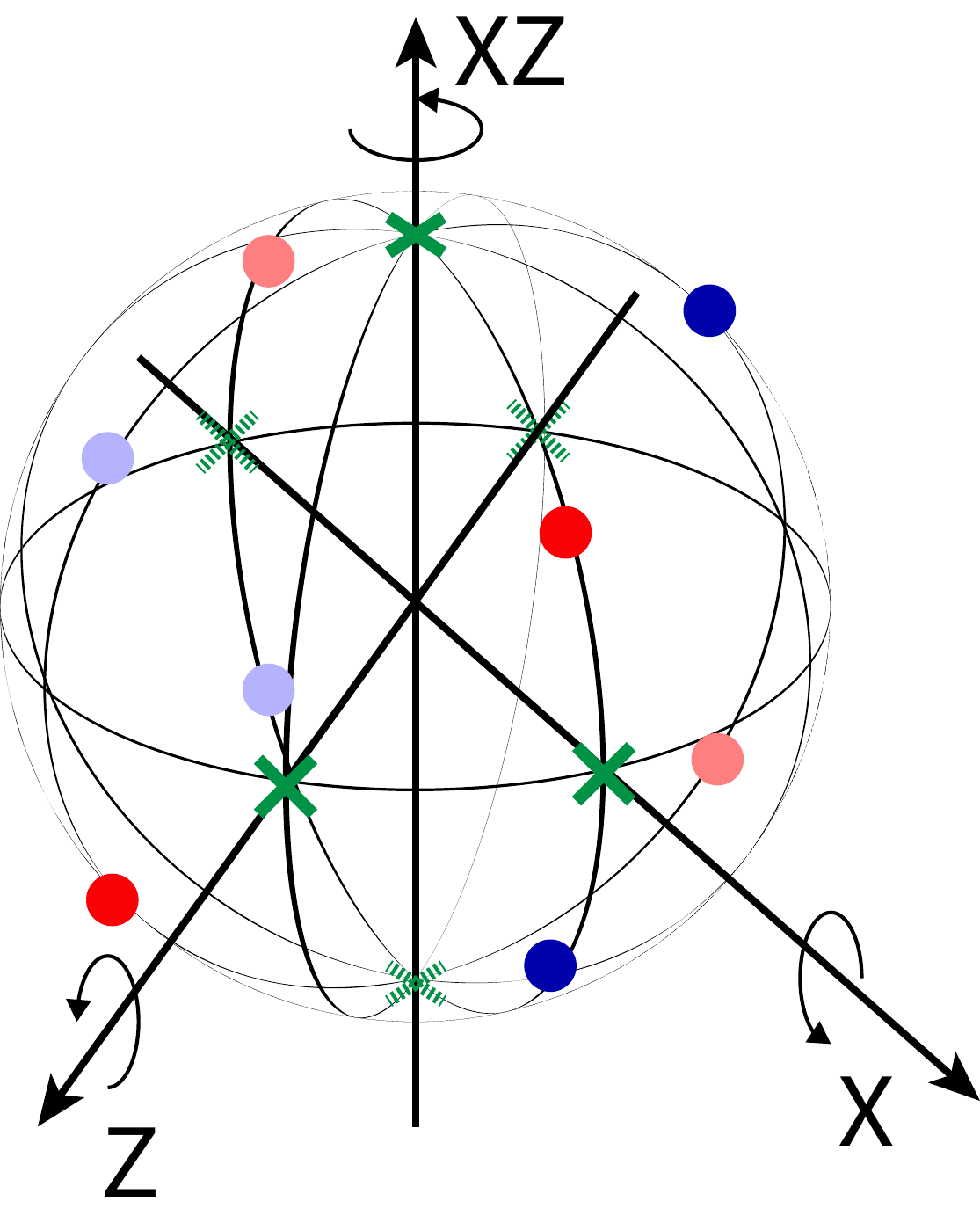}
    \caption{The constellation of the spherical code in Example \ref{example:224XZ}. The logical $0$ state is superposed by the states localized at red and pink points, while logical $1$ is by blue and light blue ones. The red and blue points have a coefficient of $1$ while the pink and light blue ones have a coefficient of $-1$. }
    \label{fig:224XZsphere}
\end{figure}

\end{example}

\begin{example}\label{example:244XZ}
The next example is the $\{2, 4, 4\}$ tessellation on the Euclidean plane. Although the symmetry $\bar{\Delta}(2, 4, 4)$ is different from the qubit Pauli group, we impose the extra relation $r_B^2=\mathbb{1}$. We obtain a two-dimensional CV code encoding a logical qubit. The logical $X$ operation is the rotation around the marked vertex along the axis perpendicular to the surface by $\pi$. The logical $Z$ and $(XZ)^{-1}$ are rotations around their corresponding vertices counterclockwise by $\frac{\pi}{2}$ respectively. In Figure \ref{fig:244XZplane}, we illustrate a specific case $(x_0,y_0)=(\frac 12, \frac 12)$. Here we take the vertex for the rotation $(XZ)^{-1}$ as the origin, horizontal direction as $x$ and vertical direction as $y$. This code can be viewed as a two-mode GKP code encodes one logical qubit. Each direction is a single-mode GKP code encoding a qubit, while the full code on the plane is a 2-to-1 concatenated qubit code using the two GKP qubits. In appendix \ref{app:244code} we describe in detail the stabilizers and logical operators from this point of view.

\begin{figure}[b]
    \centering
 \includegraphics[width=0.3\textwidth]{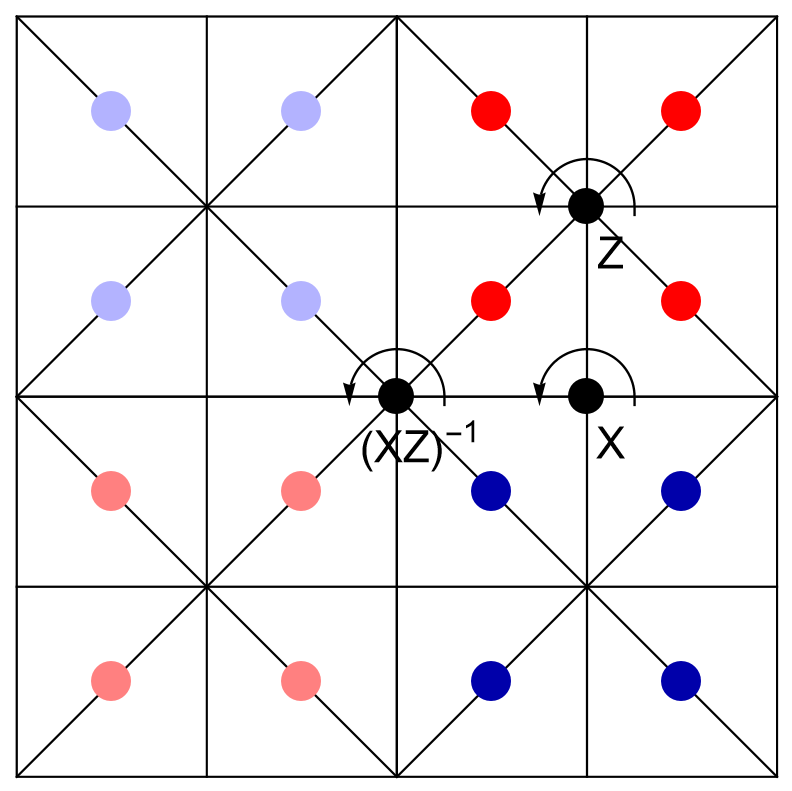}
    \caption{The unit cell of the constellation of the Euclidean plane code. The colours represent the same as those in Figure \ref{fig:224XZsphere}. The logical operations are implemented by rotation around the corresponding vertices. }
    \label{fig:244XZplane}
\end{figure}

The error correction properties of this code are analogous to those of the GKP codes. For position errors of the translation type, this code can correct any translation of distance less than the resolution $d_x=\frac{1}{2}$. The momentum errors are the plane waves. We abbreviate the momentum errors, which are plane waves, as $\hat{V}_{\Vec{k}}\equiv e^{i k_x \hat{x}} e^{i k_y \hat{y}}$. For a pair of errors $\hat{V}_{\Vec{k_1}}^\dagger\hat{V}_{\Vec{k_2}}$, write $\Delta k_x\equiv k_{x1}-k_{x2}$ and $\Delta k_y\equiv k_{y1}-k_{y2}$. The error correction condition is violated only when both $\Delta k_x, \Delta k_y$ are odd multiplicities of $\frac{\pi}{2}$. Therefore the code can correct any error with $|\Vec{k}|<\frac{\sqrt{2}\pi}{4}$. The calculation for the momentum error is in Appendix \ref{sec:244XZmomentum}.

\end{example}

\begin{example}\label{example:333XZ}
The $\{3, 3, 3\}$ tessellation can encode a logical qutrit in the plane. As shown in Table \ref{tab:codedata}, the qutrit Pauli group is obtained from the $\Bar{\Delta}(3, 3, 3)$ group by adding an extra condition indicating $\Omega=XZX^{-1}Z^{-1}$ is a central element. The logical gate $X$, $Z$ and $(ZX)^{-1}$ are implemented by rotations around the corresponding vertices by $\frac{2\pi}{3}$. Figure \ref{fig:333XZplane} shows one unit cell of the codewords with the choice $\ket{p_i}=\ket{(1,0)}$. The explicit code word is shown in Eq. \eqref{eq:honeycombcodewords}. As described in Appendix \ref{app:333code}, this code has GKP-like stabilizer operators, see Eq. \eqref{eq:333GKP}. The GKP-like logical $Z$ operator can also be obtained. However, the logical $X$ operator cannot be realized by real space displacement.

For the position error of the translation type, this code corrects translation of distance less than $d_x=\frac{\sqrt{3}}{2}.$
For the momentum error, the error correction condition is only violated when both $3\Delta k_x$ and $\sqrt{3} \Delta k_y$ are multiplicities of $\frac{2\pi}{3}$ but not multiplicities of $2\pi$. It turns out this code can correct momentum errors in any direction with $|\Vec{k}|<\frac{2\pi}{9}$. The detailed calculation is in Appendix \ref{sec:333XZmomentum}.

\begin{figure}
\centering
\includegraphics[width=0.35\textwidth]{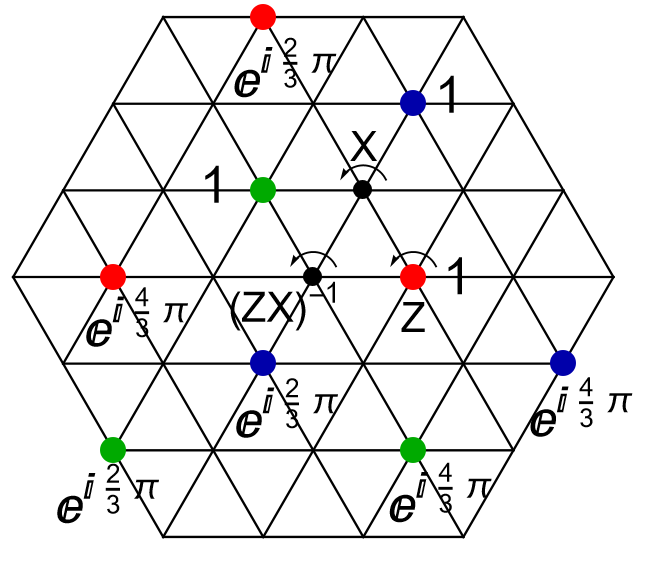}
    \caption{The unit cell of the constellation of the qutrit code in Example \ref{example:333XZ}. The red, blue and green points consist of the logical $\ket{0}$, $\ket{1}$, $\ket{2}$  states respectively. The coefficients in the superposition are labelled next to the points. The logical operations are implemented by rotation around the corresponding vertices counterclockwise by $\frac{2\pi}{3}$.}
    \label{fig:333XZplane}
\end{figure}

\end{example}

\begin{example}\label{example:555XZ}
The $\mathbb{Z}_5$ qudit Pauli group has presentation as in Eq.~\eqref{eq:quditPauli} with $d=5$. It is natural to realize it on a $\{5,5,5\}$ lattice with extra relations, as presented in Table \ref{tab:codedata}. The logical gates $X$, $Z$ and $(ZX)^{-1}$ are implemented by rotations around the corresponding vertices by $\frac{2\pi}{5}$. Figure \ref{fig:555QuditXZ} highlights one unit cell of the codewords if we choose $\ket{p_i}=\ket{(b,0)}$, where $b=\text{arccosh} \frac{\cos{\frac{\pi}{5}}+\cos^2 \frac{\pi}{5}}{\sin^2 \frac{\pi}{5}}$ is the distance between the vertices of $(ZX)^{-1}$ and $Z$. For the position error of the translation type, this code corrects translation of distance less than $d_x=\frac{1}{2} \text{arccosh}~ ((1-\cos{\frac{2\pi}{5}})\cosh^2 b+\cos{\frac{2\pi}{5}})\approx 1.6169$. Because of the $5$-fold rotation symmetry of the hyperbolic lattice,  this code can correct any momentum error $P_{-\frac 12+i r}^n(\cosh \hat{\rho})e^{i n \hat\phi}$ with $n<5$. 

Regarding the momentum error's $r$ index in hyperbolic plane, we note that the surface from the quotient $\mathbb{H}^2/\Gamma$ is a compact surface, with genus $26$ in this example. The Laplacian on this surface has a discrete spectrum, and its first excited eigenvalue $\lambda_1$ defines the energy gap. We anticipate that any error with eigenvalue less than $\lambda_1$ should satisfy the error correction condition and be correctable. However, given the generalized stabilizer data, although there is an estimate of the upper bound \cite{YangYau}, the exact value of the first positive eigenvalue has no analytic expression and it is highly non-trivial to calculate it numerically for high-genus surfaces. Nevertheless, the energy gap $\lambda_1$ provides a non-zero notion of ``distance of momentum error''. This argument applies to codes on hyperbolic \textcolor{black}{planes}, including Examples \ref{example:555XZ}, \ref{example:468Clifford} and \ref{example:4352I}.

\end{example}

\begin{example}\label{example:468Clifford}
The group presentation of the Clifford group in Eq.~\eqref{eq:quotient_singleClifford} is analogous to the proper triangle group $\bar{\Delta}\{6,4,8\}$. We can use this tessellation to construct codes with geometric logical Clifford operations. In Figure \ref{fig:468Cliffordhyperbolic} we show the codeword configurations of the logical $\ket{\bar{0}}$ and $\ket{\bar{1}}$. Here we choose $\ket{p_i}=\ket{(\frac{b}{2},0)}$, where $b=\text{arccosh} \frac{\cos{\frac{\pi}{4}}+\cos \frac{\pi}{8}\cos \frac{\pi}{6}}{\sin \frac{\pi}{8}\sin \frac{\pi}{6}}$ is the distance between the vertices of $S$ and $U$. The logical $S$ gate is realized by rotation around the corresponding vertex by $\frac{\pi}{4}$, while the logical $U$ gate is realized by rotation around the corresponding vertex by $\frac{\pi}{3}$. These gates are defined in Eq. \eqref{eq:Cliffordgates}. Interestingly, two codewords have positional overlap, but the relative phases guarantee that they have a zero inner product. This is analogous to the logical $\ket{\bar{\pm}}$ states of GKP code. For the position error of the translation type, this code corrects translation of distance less than $d_x=\frac{1}{2} \text{arccosh}~ ((1-\cos{\frac{\pi}{4}})\cosh^2 \frac{b}{2}+\cos{\frac{\pi}{4}})\approx 0.6605$.

\end{example}

\begin{example}\label{example:4352I}
We can construct a code with the logical binary icosahedral group \cite{PARZANCHEVSKI2018869, kubischta2023family} on the tessellation $\{4,3,5\}$. This group has $120$ elements and contains non-Clifford gates. It can be generated by 
\begin{equation}
   F=\frac{e^{-i\frac{\pi}{4}}}{\sqrt{2}}\begin{pmatrix}
        1&-i\\1&i\\
    \end{pmatrix},~ \Phi=\frac{1}{2}\begin{pmatrix}
        \phi_G+ i\phi_G^{-1}&1\\-1&\phi_G- i\phi_G^{-1}
    \end{pmatrix}.
\end{equation}
Here $\phi_G=\frac{\sqrt{5}+1}{2}$ is the golden ratio. The generators have the relations $F^3=\Phi^5=(\Phi F^{-1})^2=-\mathbbm{1}$. In Figure \ref{fig:468Cliffordhyperbolic} we show the codeword configurations of the logical $\ket{\bar{0}}$ and $\ket{\bar{1}}$. Here we choose $\ket{p_i}=\ket{(\frac{b}{2},0)}$, where $b=\text{arccosh} \frac{\cos{\frac{\pi}{3}}+\cos \frac{\pi}{4}\cos \frac{\pi}{5}}{\sin \frac{\pi}{4}\sin \frac{\pi}{5}}$ is the distance between the vertices of $-F$ and $-\Phi^{-1}$. The logical $-F$ gate is realized by rotation around the corresponding vertex by $\frac{2\pi}{3}$, while the logical $-\Phi^{-1}$ gate is realized by rotation around the corresponding vertex by $\frac{2\pi}{5}$. These gates are defined in Eq. \eqref{eq:Cliffordgates}. For the position error of the translation type, this code corrects translation of distance less than $d_x=\frac{1}{2} \text{arccosh}~ ((1-\cos{\frac{2\pi}{5}})\cosh^2 \frac{b}{2}+\cos{\frac{2\pi}{5}})\approx 0.5011$.
\end{example}

\section{Universal gate set on hyperbolic surfaces}\label{sec:universal}

Although the code constructions discussed above mostly focus on finite logical groups, 
we note that it is possible to realize the universal single-qubit logical gate set within our formalism. Hyperbolic \textcolor{black}{planes} admit a special family of tessellations, one or two of the labeling integers $\{p,q,r\}$ could be $\infty$. For example, the $\{\infty,2,8\}$ tessellation has presentation $r_B^2=r_C^8=\mathbbm{1}$. It is natural to identify $r_B$ with the Hadamard gate $H$ and $r_C$ with the $T$ gate. By imposing the extra relation $(r_B r_C^2)^3$ commutes with every generator and $(r_B r_C^2)^{24}=\mathbbm{1}$, the resulting quotient of the proper triangle group is isomorphic to the universal gate set. This indicates the feasibility of using Eq.~\eqref{eq:encodingmap} to construct a code whose entire logical gate set is implemented by geometric rotations. Our construction utilizes infinite-dimensional physical Hilbert spaces to circumvent the limitations on logical gates in finite dimensions. It is compatible with the analysis of codes covariant for continuous groups in Ref.~\cite{QuantumReferenceFrame}.

\section{Discussion}
In this work, we introduced a \textcolor{black}{versatile} geometric formalism for constructing continuous variable quantum error-correcting codes whose designated logical gate sets can be realized by spatial rotations, providing new perspectives for designing codes \textcolor{black}{and logical gates}. The symmetries of the tessellation lattice and the underlying surfaces play a crucial role in code construction and error correction.

Numerous directions are worth further exploring.  First, the error correction protocol warrants a more comprehensive study. Specifically, we discussed the correction for position errors, but the case of momentum errors has not been fully understood and should be further investigated.  
Second, we would like to further understand the geometric analogues of the Eastin--Knill theorem and its extensions~\cite{eastinknill,faist2020continuous,Woods_2020,zhou2021new,kubica2021using,liu2021quantum,liu2023approximate} that characterize the general interplay between \textcolor{black}{versatility of} logical operations \textcolor{black}{for a quantum code} and \textcolor{black}{its} error correction \textcolor{black}{performance}.
Third, we expect further studies of the geometric interpretation and physical relevance of magic in similar settings~\cite{hahn2022quantifying, feng2024connecting,hahn2024bridging} to be fruitful.
Fourth, the code state written formally in Eq. \eqref{eq:encodingmap} is unnormalizable for the codes on the Euclidean or hyperbolic \textcolor{black}{planes}. It is interesting to study the regularized version of the code and how exact logical gates become approximate under the regularization. Fifth, generalizing our formalism to logical gates among several encoded degrees of freedom is also an important future direction for realizing universal quantum computing. 

Furthermore, our formalism opens up new possibilities for experimental implementations of logical quantum computing through real-space geometric manipulations. Note that the lattices in curved manifolds have been implemented in experimental platforms \cite{barredo2018synthetic,kollar2019hyperbolic,lenggenhager2022simulating,chen2023hyperbolic}. The geometric rotations can be realized via combinations of beam splitters and two-mode squeezing, which are within the current scope of the experimental technology. \textcolor{black}{Our geometric logical gates have good fault tolerance properties since they do not amplify noises. First, they can be realized by two-mode Gaussian operations, which are known to be fault-tolerant for CV codes. Specifically, rotations are isometries of the surfaces, so they preserve the distances and do not amplify position errors. Rotations also commute with the Laplacian. Therefore, they preserve the eigenvalue of the momentum errors, which means a low-energy momentum error remains the same energy. As a result, we expect future exploration of the application of the tessellation codes and geometric gates in quantum experiments to be feasible and valuable.} 

\section*{acknowledgment}
We thank Victor V.~Albert, Anthony Leverrier, Jiuci Xu, Christophe Vuillot, Zhenbin Yang, Ning Bao and Ye Wang for helpful discussions. The unit cell figures on the hyperbolic planes are generated with the assistance of the GAP package \textsf{HyperCells} \cite{HyperCells}, the Mathematica package \textsf{HyperBloch} \cite{HyperBloch} and the data of the quotient of triangle groups in Ref.~\cite{Conder:2007}. Y.W. is supported by the National Natural Science Foundation of China Grant No. 12347173, China Postdoctoral Science Foundation Grant No. 2023M742003, and the Shuimu Tsinghua Scholar Program of Tsinghua University. Z.-W.L.\ is supported in part by a startup funding from YMSC, Tsinghua University, and NSFC under Grant No.~12475023.

\newpage

\begin{widetext}

\begin{appendix}
\section{Distances on surfaces and the resolution}\label{app:constellation}
In this section, we present the formula for calculating the distances of two points on the sphere and the hyperbolic plane.

For the two-dimensional sphere $S^2$, we conventionally embed it into the three-dimensional flat space $\mathbb{R}^{3}$:
\begin{eqs}\label{eq:sphericalembedding}
&X_1=\sin{\theta}\cos\phi,~X_2=\sin{\theta}\sin\phi,~X_3=\cos{\theta},\quad X_1^2+X_2^2+X_3^2=1,~ ds^2=dX_1^2+dX_2^2+dX_3^2.
\end{eqs}
Because the sphere is a homogeneous and isotropic space, to calculate the distance of two arbitrary points on the sphere, we can always find an isometry that transforms one point to $p_1=(\theta=0,\phi=0)$ and another to a certain $p_2=(\theta=\theta_0,\phi=0)$. In this case, it is not difficult to calculate the geodesic distance between them as $\theta_0$. Note that if we write the res in the embedding coordinate, we have 
\begin{equation}\label{eq:spheredistance}
    d(p_1,~p_2)=\arccos \Vec{X}^{(1)}\cdot \Vec{X}^{(2)},
\end{equation}
where $\Vec{X}^{(i)}=(\sin{\theta}\cos\phi,\sin{\theta}\sin\phi,\cos\theta)|_{p_i}$. Because the inner product is invariant under any isometry of the sphere, the above expression applies to arbitrary two points on the sphere.

For the hyperbolic surface, 
it is also convenient to embed it into $\mathbb{R}^{2,1}$.
\begin{eqs}\label{eq:hyperboilcembedding}
&X_0=\cosh{\eta},~X_1=\sinh{\eta}\cos\theta,~X_2=\sinh{\eta}\sin\theta, \quad -X_0^2+X_1^2+X_2^2=-1,~ ds^2=-dX_0^2+dX_1^2+dX_2^2.
\end{eqs}
The resulting line element is
$ds^2=d\eta^2+\sinh^2{\eta} d\theta^2$. Because the hyperbolic space is also homogeneous and isotropic, we adopt the same strategy as in the sphere case to calculate the distance between two points. We first transform two points such that $p_1$ is at the origin, $p_2$ at $(\eta_0,\theta=0)$. It is easy to calculate their distance in this case as $\eta_0$. Writing in terms of the embedding coordinate, it is 
\begin{equation}\label{eq:hyperbolicdistance}
d(p_1,~p_2)=\text{arccosh}~(-\Vec{X}^{(1)}\cdot \Vec{X}^{(2)}),
\end{equation}
where $\Vec{X}^{(i)}=(\cosh{\eta},\sinh{\eta}\cos\theta,\sinh{\eta}\sin\theta)|_{p_i}$ and $-\Vec{X}^{(1)}\cdot \Vec{X}^{(2)}=X_0^{(1)}X_0^{(2)}-X_1^{(1)}X_1^{(2)}-X_1^{(1)}X_1^{(2)}$. Similar to the spherical case, because the inner product is invariant under isometry, this equation applies to any two points in the hyperbolic surface.

One special feature of the hyperbolic plane is that given the values of the three angles $A,B,C$ of a triangle, its sides are unambiguously determined. This is not the case for the sphere or the Euclidean plane. Let the length of the sides opposite the the corresponding angles be $a,b,c$, we have
\begin{eqs}
\cosh{a}=\frac{\cos A+\cos B \cos C}{\sin{B}\sin{C}},~
\cosh{b}=\frac{\cos{B}+\cos{C}\cos A}{\sin{C}\sin{A}},~\cosh{c}=\frac{\cos{C}+\cos{A}\cos B}{\sin{A}\sin{B}}.
\end{eqs}

To calculate the resolution, which is one-half of the minimal distance among any pair of points, we need to get an expression of the distance between points before and after rotating around a point. This can be calculated using Eq. \eqref{eq:spheredistance} and \eqref{eq:hyperbolicdistance}. For a point $p_0$ on the sphere, if rotated around a point $j$ by angle $\alpha$ to get $p_{0j}$, then $\Vec{X}^{(p_0)}\cdot \Vec{X}^{(p_{0j})}=\sin^2 d_{0j}\cos\alpha+\cos^2 d_{0j}$, where $d_{0j}=d(p_0,p_j)$, the distance between $p_0$ and $p_j$. A similar expression in the hyperbolic case is $\Vec{X}^{(p_0)}\cdot \Vec{X}^{(p_{0j})}=\cosh^2 d_{0j}-\sinh^2 d_{0j}\cos\alpha$. To obtain the resolution, it is enough to evaluate the distances of the point $p_i$ in Eq.~\eqref{eq:encodingmap} with respect to its rotated points after the rotation around $A,B,C$. Therefore, 
\begin{eqs}
    d_x=\frac{1}{2} \min\{d(p_i, p_A), d(p_i, p_B), d(p_i, p_C)\}.
\end{eqs}
Here $A,B,C$ are the vertices of the triangle and are usually identified with the rotation vertices of the logical operations in specific code constructions. 

For a fixed tessellation, one may optimize the choice of the state $\ket{i}$ in the code construction to maximize the resolution. This is achieved by solving the optimization problem 
\begin{eqs}\label{eq:optimalresolution}
    p_i^{(o)}=\underset{p_i }{\text{argmax}} \min\{d(p_i, p_A), d(p_i, p_B), d(p_i, p_C)\}.
\end{eqs}
For the cases of our interest, the optimal $ p_i^{(o)}$ is a solution of $d(p_i^{(o)}, p_A)= d(p_i^{(o)}, p_B)=d(p_i^{(o)}, p_C)$.

\textcolor{black}{\section{Encoding map and codewords}}

\textcolor{black}{In this section, we provide various detailed information on the construction of our tessellation codes. We begin with an explicit calculation of the covariant encoding map and verify its operator pushing property, which ensures that the encoding is covariant under group $G$ and thus carries the desired logical operations. Then, we describe the systematic and versatile framework for generating tessellations codes in the main text, and provide explicit constructions of the codewords for each example.}

\subsection{Covariant encoding map}
We begin by demonstrating the operator pushing identity $\rho(h) W= W \rho_L(h)$, for all $ h\in G$:
\begin{eqs}
    \rho(h) W = \sum_{g \in G} \rho_L(g)^{-1} \otimes \rho (hg) = \sum_{g \in G} [\rho_L(h^{-1}g)]^{-1} \otimes \rho(g)= \sum_{g \in G} \rho_L(g)^{-1}\rho_L(h) \otimes \rho(g)=W \rho_L(h).
\end{eqs}
The encoding $\mathcal{E}(\ket{k})$ can be expressed as an isometric operator $V_G$ acting on the input logical state $\ket{k}$, where the isometry is given by
\begin{eqs}
    V_G= \bra{\Sigma}W \ket{\text{init}}=\sum_{g \in G} \bra{\Sigma}\rho_L(g)^{-1} \otimes \rho(g) \ket{\text{init}}.
\end{eqs}
By choosing $\rho(g)=\pi(g)$, $\rho_L(g)^{-1}=\lambda(g)^{-1}$, $\bra{\Sigma}=\bra{\Omega}$, $ \ket{\text{init}}=\ket{\Phi}$ in $V_G$, the isometry $V_G$ reduces to the form of the encoding isometry presented in Ref.~\cite{denys2023multimode}.

\textcolor{black}{Finally, the encoding map can be written as 
\begin{eqs}
    \mathcal{E}(\ket{k})\equiv \ket{\overline{k}}\propto V_G \ket{k}= \sum_{ g \in G} \bra{\Sigma} \rho_L(g)^{-1} \ket{k} \rho(g) \ket{\text{init}},
\end{eqs}
where the encoding isometry $V_G$ maps the input logical state $\ket{k}$ to the corresponding codeword $\ket{\overline{k}}$.}

\subsection{Systematic framework for tessellation code construction} \label{app:code_construction}
We now clarify the systematic framework for constructing tessellation codes with geometric logical gates.
At a high level, the starting point is the choice of a logical group $G$, with the goal of building a code that is covariant under its action. The key is to identify a tessellation whose symmetry structure is compatible with that of the desired logical group $G$. Specifically, the general procedure goes as follows:
\begin{enumerate}
    \item \textbf{Choose a logical group $G$.}\\
    Analyze the group structure of $G$ in terms of its group presentation.  We desire to construct a code that is covariant with respect to $G$. 
In particular, we require that $G$ admits a presentation of the form
\begin{equation}
    G=\langle g_1, g_2| g_1^p=g_2^q=(g_1 g_2)^r=\mathbbm{1}, r_1=...=\mathbbm{1}\rangle,
\end{equation}
where the relations $r_i$ are the possible extra relations corresponding to generalized stabilizers.
\item \textbf{Identify a compatible tessellation.} \\
We select a tessellation specified by the triple $\{t_1,t_2,t_3\}$, corresponding to a triangle group with presentation
\begin{equation}
    \bar\Delta=\langle r_A, r_B| r_A^{t_1}=r_B^{t_2}=(r_A r_B)^{t_3}=\mathbbm{1}\rangle.
\end{equation}
The natural compatibility condition is $\{t_1,t_2,t_3\}=\{p,q,r\}$. The type of plane to implement the codewords is also determined by the type of tessellation, via Eq.~(7) in the manuscript. The generators are also identified $r_A \leftrightarrow g_1$, $r_B \leftrightarrow g_2$. Using this identification between generators in $G$ and $\bar\Delta$, the extra relations $\{r_1,...\}$ in $G$ are mapped to $\{s_1,...\}$ in $\bar{\Delta}$, which serve as the generalized stabilizers.
\item \textbf{Construct the encoding map and calculate the codewords.}\\
The encoding map, given in Eq.~(9) in the main text, is expressed as a superposition of delta functions with corresponding coefficients. The two types of elements are obtained as follows:
\begin{enumerate}
    \item Calculate the coefficients $\bra{\Sigma}\rho_L(g)^\dagger \ket{k}$ which represent the overlap between the logical state $\ket{k}$, acted by the logical representation $\rho_L(g)^\dagger $, and the reference state $\ket{\Sigma}$.
    \item Evaluate the geometric action $\rho(g)\ket{\text{inti}}$ by decomposing the group element $g$ into a product of generators, and then sequentially applying the geometric action of each generator in the order specified by this decomposition, and summing over all stabilizer-equivalent form of $g$.
\end{enumerate}
\end{enumerate}

\subsection{Computation of codewords}

We summarize the general procedure for code construction described in Sec.~\ref{app:code_construction} as the combination of the following data:
\begin{enumerate}
    \item The unitary representations of the logical group generators within the logical subspace, i.e., the unitary matrices corresponding to their action on the logical space, which can give us $\bra{\Sigma} \rho_L(g)^\dagger \ket{k}$.
    \item The decomposition of logical group elements as products of generators.
    \item The geometric action of each logical group generator on the tessellated planes, which gives $\rho(g) \ket{\text{init}}$.
\end{enumerate}
With the above information, we can enumerate $g$ over every group element, calculate the coefficients $\bra{\Sigma}\rho_L(g)^\dagger\ket{k}$ and the geometric action $\rho(g)\ket{\text{inti}}$, and then superpose all the branches to obtain the codeword $\sum_{g\in G}\bra{\Sigma} \rho_L(g)^\dagger\ket{k}\rho(g) \ket{\text{init}}$.
In this subsection, we calculate the codewords of all specific tessellation code examples in this paper following the procedure outlined above. We provide the necessary data for hyperbolic examples without explicitly writing out the resulting codewords, because the superpositions in each unit cell are too lengthy, and summing over all unit cells does not have a nice closed form as in Euclidean cases.

\subsubsection{Example \ref{example:224XZ}}\label{app:224code}

For Example \ref{example:224XZ}, the logical group is the qubit Pauli group. The logical representations of the generators are the Pauli $X$ and $Z$ matrices
\begin{eqs}
\mathcal{P}_{\text{qubit}}=\langle X,Z| X^2=Z^2=(XZ)^4=\mathbbm{1}\rangle,
~\rho_L(X)=\begin{pmatrix}
        0&1\\1&0\\
    \end{pmatrix},~
\rho_L(Z)=\begin{pmatrix}
        1&0\\0&-1\\
    \end{pmatrix}.   
\end{eqs}
The qubit Pauli group $\mathcal{P}_{\text{qubit}}$  has $8$ elements, which can be written in terms of generators as
\begin{eqs}
    \mathcal{P}_{\text{qubit}}=\{Z^i(XZ)^j |\forall i=0,1, j=0,1,2,3\}.
    \end{eqs}

The compatible tessellation is $\{2,2,4\}$, which lives on a sphere. Its proper triangle group is $\bar{\Delta}(2,2,4)=\langle r_A, r_B|~r_A^2=r_B^2=(r_A r_B)^4=\mathbbm{1}\rangle$. The group generators are identified as $Z \rightarrow r_A, X \rightarrow r_B, XZ \rightarrow r_B r_A$. Therefore, the physical representations of the Pauli group generators are $\rho(Z)=r_A, \rho(X)=r_B$. To calculate the action of $r_A$ and $r_B$ on the sphere, we use the standard embedding of a two-dimensional sphere into three-dimensional flat space as shown in (\ref{eq:sphericalembedding}). In this embedding, $r_A$ is rotation around the $x$ axis by $\pi$, $r_B$ is rotation around the bisector of the positive $x$ and $y$ axes by $\pi$, see Figure \ref{fig:224XZsphere2} for an illustration. Therefore, they take the form 
\begin{eqs}\label{eq:sphericalgeneratorphysicalrep}
    r_A=\rho(Z)=\begin{pmatrix}
        1&0&0\\0&-1&0\\0&0&-1
    \end{pmatrix},\quad r_B=\rho(X)=\begin{pmatrix}
        0&1&0\\1&0&0\\0&0&-1
    \end{pmatrix}.
\end{eqs}

\textcolor{black}{We pick a point on a two-dimensional sphere $\mathbb{S}^2$ as our initial state $\ket{\text{init}}=$
 $\ket{p_i}=\ket{\theta_0, \phi_0}$ in the construction Eq.~\eqref{eq:encodingmap} and set $\ket{\Sigma}=\ket{0}$. Then we can write the general form of the codewords as
\begin{eqs}\label{eq:224_encoding_map}
    \ket{\overline{0}}&\propto \sum_{g\in \mathcal{P}_{\text{qubit}}} \bra{0} \rho_L(g)^\dagger \ket{0} \rho(g) \ket{p_i}\\
    &=\bra{0} I \ket{0}  \ket{p_i} +  \bra{0} (ZX)^2 \ket{0} (r_B r_A)^2 \ket{p_i}+ \bra{0} Z \ket{0} r_A \ket{p_i}+ \bra{0} (ZX)^2 Z \ket{0} r_A (r_B r_A)^2 \ket{p_i}\\
    &= \ket{p_i} - (r_B r_A)^2 \ket{p_i} + r_A \ket{p_i} - r_A (r_B r_A)^2 \ket{p_i}\\
    \ket{\overline{1}}&\propto \sum_{g\in \mathcal{P}_{\text{qubit}}} \bra{0} \rho_L(g)^\dagger \ket{1} \rho(g) \ket{p_i}\\
    &=\bra{0} X \ket{1} r_B \ket{p_i} + \bra{0} -X \ket{1} (r_B r_A)^2 r_B \ket{p_i} +\bra{0} ZX \ket{1} r_B r_A \ket{p_i}+ \bra{0} (ZX)^3 \ket{1} (r_B r_A)^3 \ket{p_i}\\
    &=r_B \ket{p_i} - (r_B r_A)^2 r_B \ket{p_i}+ r_B r_A \ket{p_i}- (r_B r_A)^3 \ket{p_i}.
\end{eqs}
Here, each codeword involves only $4$ terms, as all other terms have zero coefficients. The action of $r_A$ and $r_B$ on $\ket{p_i}$ is calculated by using the spherical coordinate $\vec{p_i}=(\sin\theta_0\cos\phi_0,\sin\theta_0\sin\phi_0,\cos\theta_0)^T$, and acting on the matrix representation in (\ref{eq:sphericalgeneratorphysicalrep}) from the left. This yields codewords} 
\begin{eqs}\label{eq:224_codewords}
    \ket{\overline{0}}\propto &\ket{\theta_0, \phi_0}+\ket{\pi-\theta_0,-\phi_0}  -\ket{\theta_0, \pi+\phi_0}-\ket{\pi-\theta_0,\pi-\phi_0},\\
    \ket{\overline{1}}\propto &\ket{\theta_0, \phi_0+\frac{\pi}{2}}+\ket{\pi-\theta_0,-\phi_0+\frac{\pi}{2}} -\ket{\theta_0, -\frac{\pi}{2}+\phi_0}-\ket{\pi-\theta_0,-\frac{\pi}{2}-\phi_0}.
\end{eqs}

In Example \ref{example:224XZ}, the initial state is chosen as $(\theta_0=\arccos{\frac{1}{\sqrt{3}}},\phi_0=\frac{\pi}{4})$. Here we show two other cases. They are depicted in Figure \ref{fig:224XZsphere2}.

\begin{enumerate}
\item \label{item:224XZflat} \textbf{Case 1:} Another natural choice is $\theta_0= \frac{\pi}{2}$, $\phi_0=0$. The codeword configuration is illustrated in the panel (b) of Fig. \ref{fig:224XZsphere2}. The logical operations are implemented in the same manner as in Example \ref{example:224XZ}. The resolution, which is half of the minimal distance between the configuration points is $\frac{\pi}{4}$. Therefore, this code can correct any rotation with a rotation angle less than $\frac{\pi}{4}$. However, the error correction condition will be violated for those rotations whose axis passes through the states' configuration points, unlike the code in Example \ref{example:224XZ}. For momentum errors, we use Eq. \eqref{eq:sphericalmomentumerror} with $\theta_0= \frac{\pi}{2}$, $\phi_0=0$. It is straightforward to find that the lowest error pair that violates the error correction condition is $Y_0^{0\dagger}Y_2^{\pm 2}$. So this code can also correct any error with $\ell\leq 1$.

    \item \label{item:224XZoptimal} \textbf{Case 2:} If we exclude the cases in which $\ket{p_i}$ is put on the vertices of rotation, we can solve for the maximal resolution choice of $\ket{p_i}$ of Eq. \ref{eq:optimalresolution}. The solution is $\theta_0=\arccos{\sqrt{\frac{\sqrt{2}}{4+\sqrt{2}}}},~\phi_0=\frac{\pi}{8}$. We depict the codeword configuration of this choice of $\ket{p_i}$ in the right panel of Figure \ref{fig:224XZsphere2}. The resolution turns out to be $d_x=\frac{1}{2}\arccos{\frac{\sqrt{2}}{4+\sqrt{2}}}\approx 0.6533$, which is slightly larger than the case in Example \ref{example:224XZ}, where $d_x=\frac{1}{2}\arccos{\frac{1}{3}}\approx 0.6155$. This code corrects any rotation error whose rotation angle is less than the resolution $d_x\approx 0.6533$. But similar to Case \ref{item:224XZflat}, it does not correct the errors when the rotation axis of $R_2^\dagger R_1$ passes through any codeword configuration point. For the momentum error, plugging $\theta_0=\arccos{\sqrt{\frac{\sqrt{2}}{4+\sqrt{2}}}},~\phi_0=\frac{\pi}{8}$ into Eq. \eqref{eq:224_detection_condition}, we see that the error correction condition is violated when $(m_2-m_1) \mod 4=2$. 
\end{enumerate}

\begin{figure}
\centering
    \subfloat[$(\theta_0=\arccos{\frac{1}{\sqrt{3}}},\phi_0=\frac{\pi}{4})$]{\includegraphics[width=0.27\textwidth]{224_spherical.pdf}}
     \hfill
     \subfloat[$(\theta_0=\frac{\pi}{2},\phi_0=0)$]{\includegraphics[width=0.27\textwidth]{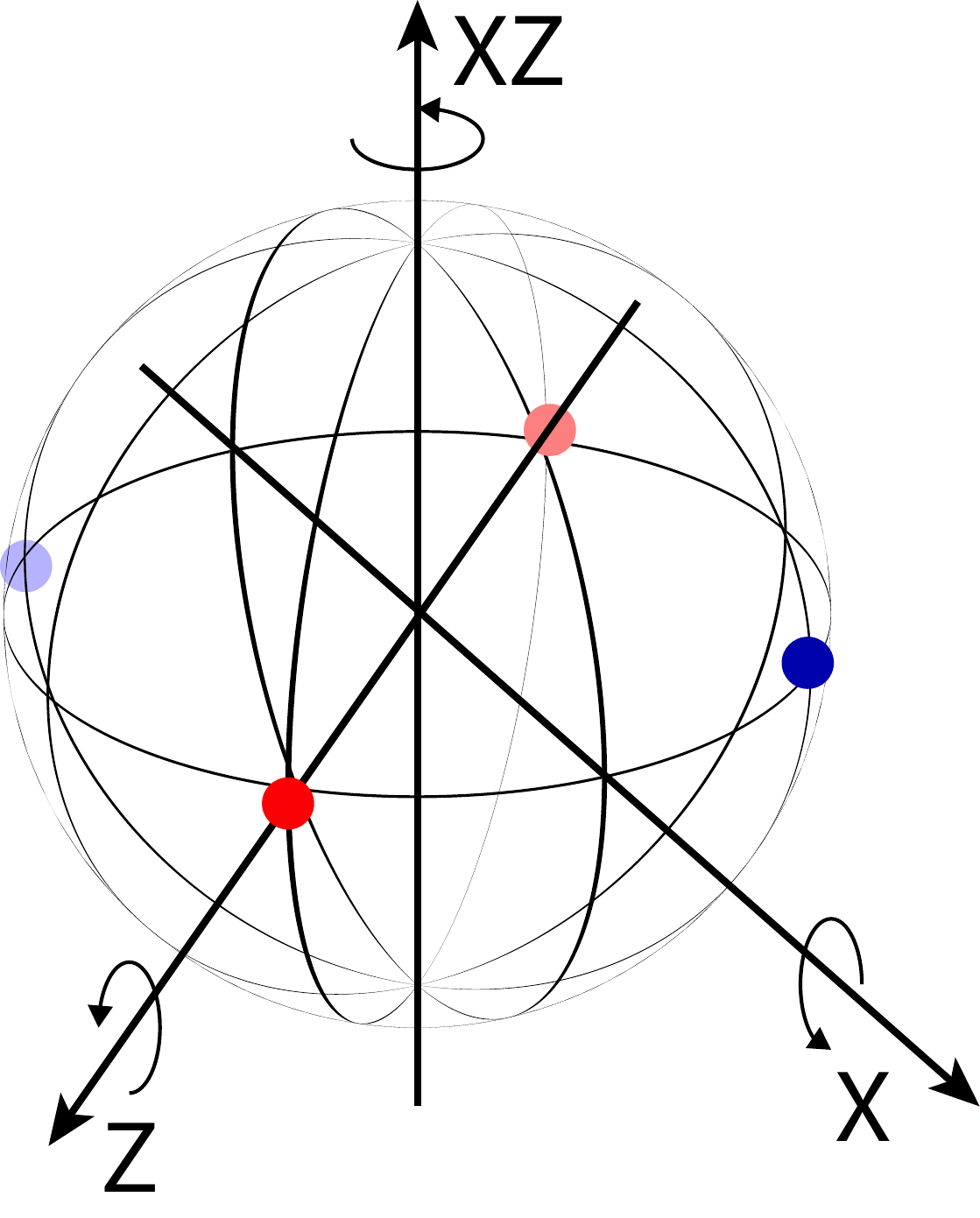}}
     \hfill
     \subfloat[$(\theta_0=\arccos{\sqrt{\frac{\sqrt{2}}{4+\sqrt{2}}}},\phi_0=\frac{\pi}{8})$]{\includegraphics[width=0.27\textwidth]{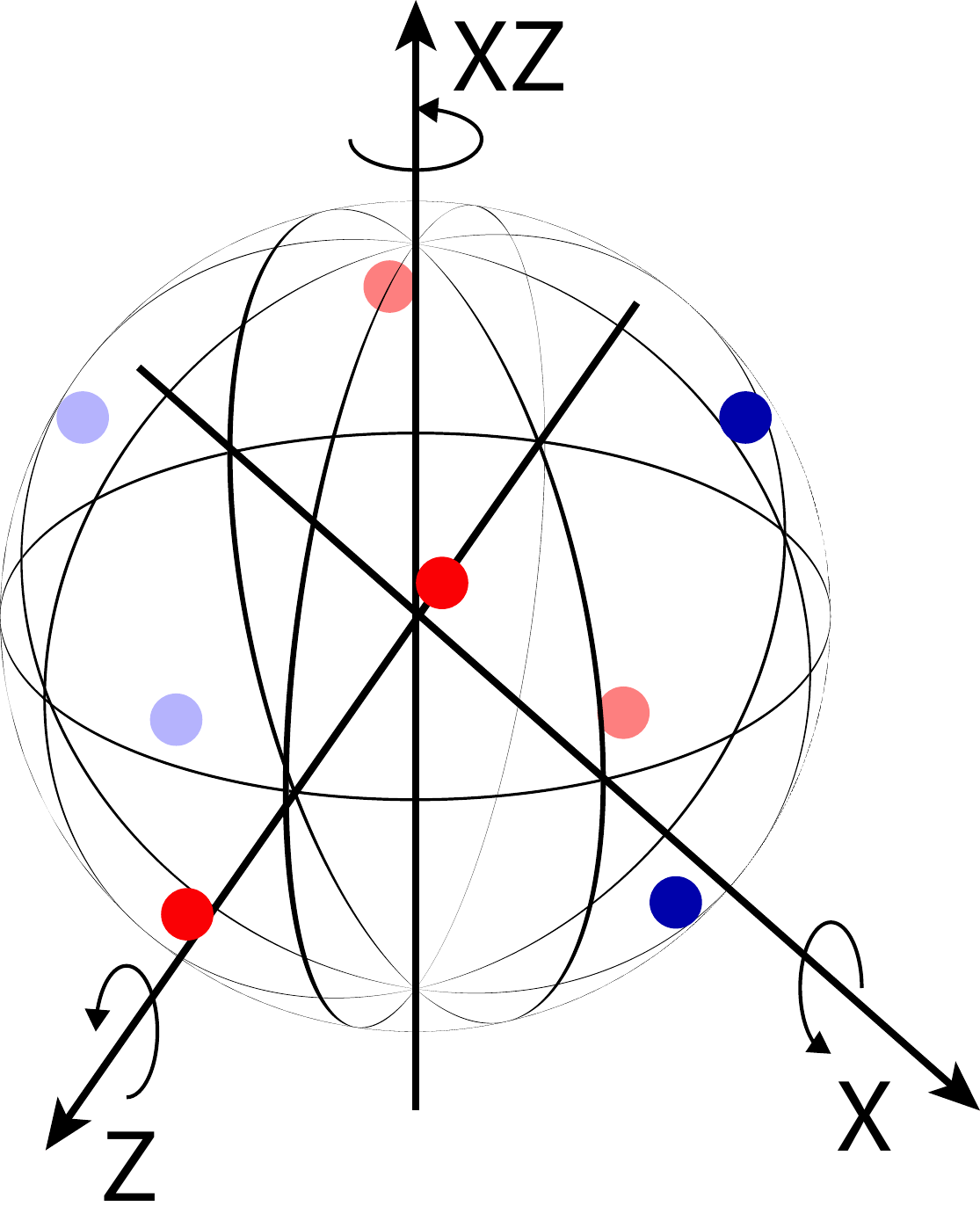}}
     \hfill
    \caption{This figure compares the code constructed from the same encoding map but with a different initial state $\ket{p_i}$. The configurations of the spherical code in Example \ref{example:224XZ}, Case~\ref{item:224XZflat} and Case~\ref{item:224XZoptimal} are shown in subfigures (a), (b) and (c) respectively. The coloring convention is the same as the one used in the main article.}
    \label{fig:224XZsphere2}
\end{figure}

\subsubsection{Example \ref{example:244XZ}}\label{app:244code}

For the code construction in Figure \ref{example:244XZ}, if we take the $(XZ)^{-1}$ rotation vertex as $(0,0)$, for a generic $\ket{p_1}=\ket{(x_0, y_0)}$, the unnormalized codewords are 
\begin{eqs}\label{eq:Euclideantessellationcodeword}
    \ket{\bar 0}=&\sum_{m,n\in \mathbb{Z}} \ket{(4m+x_0,4n+y_0)}+\ket{(4m+2-y_0,4n+x_0)}+\ket{(4m+y_0,4n+2-x_0)}+\ket{(4m+2-x_0,4n+2-y_0)}\\
    -&\ket{(4m-x_0,4n-y_0)}-\ket{(4m-2+y_0,4n-x_0)}-\ket{(4m-y_0,4n-2+x_0)}-\ket{(4m-2+x_0,4n-2+y_0)},\\
    \ket{\bar 1}=&\sum_{m,n\in \mathbb{Z}} \ket{(4m+x_0,4n+y_0-2)}+\ket{(4m+2-y_0,4n+x_0-2)}+\ket{(4m+y_0,4n-x_0)}+\ket{(4m+2-x_0,4n-y_0)}\\
   -&\ket{(4m-x_0,4n-y_0+2)}-\ket{(4m-2+y_0,4n-x_0+2)}-\ket{(4m-y_0,4n+x_0)}-\ket{(4m-2+x_0,4n+y_0)}.\\
\end{eqs}

As mentioned in the main text, each mode in the $x$ or $y$ direction is a one-dimensional GKP code. Its code words are
\begin{eqs}
\ket{k}_{\text{GKP}}&=\sum_{n\in\mathbb{Z}} \ket{4n+2k+\frac{1}{2}}+\ket{4n+2k+\frac{3}{2}},~k\in\{0,1\}.\\
\end{eqs}
The GKP stabilizers for each mode are $\hat{S}_q=e^{i 2\pi(\hat{q}-\frac{1}{2})}$ and $\hat{S}_p=e^{i4\hat{p}}$. The GKP logical operators are $X_{\text{GKP}}=e^{i2\hat{p}},~Z_{\text{GKP}}=\sqrt{2}\sin \frac{\pi}{2}\hat{q}$. The codewords in Eq.~\eqref{eq:Euclideantessellationcodeword} can be written as 
\begin{eqs}
    \ket{\bar{0}}=\frac{1}{\sqrt{2}}(\ket{0}_{\text{GKP}}^{(x)}\ket{0}_{\text{GKP}}^{(y)}-\ket{1}_{\text{GKP}}^{(x)}\ket{1}_{\text{GKP}}^{(y)}),~
     \ket{\bar{1}}&=\frac{1}{\sqrt{2}}(\ket{0}_{\text{GKP}}^{(x)}\ket{1}_{\text{GKP}}^{(y)}-\ket{1}_{\text{GKP}}^{(x)}\ket{0}_{\text{GKP}}^{(y)}).\\
\end{eqs}
This is a $2$ -to-$1$ concatenated GKP code with stabilizers\ and logical operators
\begin{equation}
S=-X_{\text{GKP}}^{(x)}X_{\text{GKP}}^{(y)},~L_X=-X_{\text{GKP}}^{(x)}=X_{\text{GKP}}^{(y)},~L_Z=Z_{\text{GKP}}^{(x)}Z_{\text{GKP}}^{(y)}.
\end{equation}

\subsubsection{Example \ref{example:333XZ}}\label{app:333code}
\textcolor{black}{For Example \ref{example:333XZ}, the covariant subgroup is the qutrit Pauli group. The logical representation of the generators is the Pauli $X$ and $Z$ matrices for qutrit.
\begin{eqs}
    \mathcal{P}_{\text{qutrit}}= \langle X,Z |X^3=Z^3=\Omega X \Omega^{-1}X^{-1}=\Omega Z \Omega^{-1}Z^{-1}=\mathbbm{1}\rangle,~ \rho_L(Z)=\begin{pmatrix}
        1&0&0\\0&e^{i\frac{2\pi}{3}}&0\\0&0&e^{i\frac{4\pi}{3}}
    \end{pmatrix},~\rho_L(X)=\begin{pmatrix}
        0&1&0\\0&0&1\\1&0&0
    \end{pmatrix}.
\end{eqs}
Here $\Omega\equiv XZX^{-1}Z^{-1}$ is the group commutator of $X$ and $Z$. $\mathcal{P}_{\text{qutrit}}$ has $27$ elements. In terms of generators, they can be written as 
\begin{eqs}\label{eq:qutritelement}
   \mathcal{P}_{\text{qutrit}}= \{\Omega^i X^j Z^k|\forall i,j,k\in\mathbb{Z}_3\} 
\end{eqs}}

The compatible tessellation is $\{3,3,3\}$ on a Euclidean plane. Its proper triangle group is $\bar{\Delta}(3,3,3)=\langle r_A, r_B|~r_A^3=r_B^3=(r_A r_B)^3=\mathbbm{1}\rangle$. The group generators are $Z \rightarrow r_A, X \rightarrow r_B, XZ \rightarrow r_B r_A$, so the physical representation of the Pauli group generators is $\rho(Z)=r_A, \rho(X)=r_B$.
If we write the initial state as $\ket{\text{init}}=\ket{(x_0,y_0)}$, $r_A$ is the rotation counterclockwise around point $(1,0)$ by $\frac{2\pi}{3}$, while $r_B$ is the rotation counterclockwise around point $(\frac{1}{2},\frac{\sqrt{3}}{2})$ by $\frac{2\pi}{3}$. The action of $r_A$ and $r_B$ is decomposed into first shifting to the origin, then performing the rotation, and finally shifting back to the starting point. For the convenience of calculation, we adopt the embedding of the two-dimensional Euclidean plane into a three-dimensional space as $\vec p=(x_0,y_0,1)^T$, so translations can be expressed as matrix multiplication. The generators take the representation 
\begin{eqs}
     r_A&=\begin{pmatrix}
        1&0&1\\0&1&0\\0&0&1
    \end{pmatrix}\begin{pmatrix}
        \cos\frac{2\pi}{3}&-\sin\frac{2\pi}{3}&0\\ \sin\frac{2\pi}{3}&\cos\frac{2\pi}{3}&0\\0&0&1
    \end{pmatrix}\begin{pmatrix}
        1&0&-1\\0&1&0\\0&0&1
    \end{pmatrix}=\begin{pmatrix}
        \cos\frac{2\pi}{3}&-\sin\frac{2\pi}{3}&\frac{3}{2}\\ \sin\frac{2\pi}{3}&\cos\frac{2\pi}{3}&-\frac{\sqrt3}{2}\\0&0&1
    \end{pmatrix},\\
    r_B&=\begin{pmatrix}
        1&0&\frac12\\0&1&\frac{\sqrt3}{2}\\0&0&1
    \end{pmatrix}\begin{pmatrix}
        \cos\frac{2\pi}{3}&-\sin\frac{2\pi}{3}&0\\ \sin\frac{2\pi}{3}&\cos\frac{2\pi}{3}&0\\0&0&1
    \end{pmatrix}\begin{pmatrix}
        1&0&-\frac12\\0&1&-\frac{\sqrt3}{2}\\0&0&1
    \end{pmatrix}=\begin{pmatrix}
        \cos\frac{2\pi}{3}&-\sin\frac{2\pi}{3}&\frac{3}{2}\\ \sin\frac{2\pi}{3}&\cos\frac{2\pi}{3}&\frac{\sqrt3}{2}\\0&0&1
    \end{pmatrix}.
\end{eqs}

Using the representation above, the generalized stabilizer $\Omega r_A \Omega^{-1} r_A^{-1}=\mathbbm{1}$ and the relation $\Omega^3=\mathbbm{1}$ generate the translation symmetry of the codewords, which are $\Vec{v}_1=(0,{3\sqrt{3}})$ and $\Vec{v}_2=(\frac{9}{2},\frac{3\sqrt{3}}{2})$. In Example \ref{example:333XZ}, we choose the initial state as $\ket{(1,0)}$. Enumerating the group elements in (\ref{eq:qutritelement}) by calculating $\bra{0}\rho_L(g)^\dagger \ket{k} \rho(g)\ket{\text{init}}$ and applying the translational stabilizer, the codewords can be written as

\begin{eqs}\label{eq:honeycombcodewords}
\ket{\bar{0}}=&\sum_{m,n\in\mathbb{Z}} T_{\vec{v}_1}^m T_{\vec{v}_2}^n\left(\ket{(1,0)}+\omega \ket{(-\frac 12,\frac{3\sqrt{3}}{2})}+\omega^2\ket{(-2,0)}\right),\\
\ket{\bar{1}}=&\sum_{m,n\in\mathbb{Z}} T_{\vec{v}_1}^m T_{\vec{v}_2}^n\left(\ket{(1,\sqrt{3})}+\omega \ket{(-\frac 12,-\frac{\sqrt{3}}{2})}+\omega^2\ket{(\frac{5}{2},-\frac{\sqrt{3}}{2})}\right),\\
\ket{\bar{2}}=&\sum_{m,n\in\mathbb{Z}} T_{\vec{v}_1}^m T_{\vec{v}_2}^n\left(\ket{(-\frac{1}{2},\frac{\sqrt{3}}{2})}+\omega \ket{(-2,-\sqrt{3})}+\omega^2\ket{(1,-\sqrt{3})}\right),
\end{eqs}
where $\omega=e^{i \frac{2\pi}{3}}$.
Here $T_{\vec{v}_1}^m T_{\vec{v}_2}^n$ is a translation operator that translate a quantum state $\ket{\vec{u}}$ to $\ket{\vec{u}+m\vec{v}_1 + n \vec{v}_2}$. This code has GKP-like stabilizer operators.  The stabilizers are
\begin{eqs}\label{eq:333GKP}
    &\hat{S}_{p,1}=e^{i 3\sqrt{3}\hat{p}_y},~  \hat{S}_{p,2}=e^{i (\frac{9}{2}\hat{p}_x+\frac{3\sqrt{3}}{2}\hat{p}_y)},~\hat{S}_{q,1}=e^{i \frac{4\pi}{\sqrt{3}}\hat y},~ \hat{S}_{q,1}=e^{i (\frac{2\pi}{3}(\hat x-1)+\frac{2\pi}{\sqrt{3}}\hat y)}.
\end{eqs}
Apart from the rotational logical operators, this code has a GKP-like logical $Z$ operator, which is $L_Z=e^{i\frac{4\pi}{3\sqrt{3}}\hat{y}}$ and its stabilizer equivalence. However, the logical $X$ operator cannot be realized by real space displacement.

\textcolor{black}{\subsubsection{Example \ref{example:555XZ}}}
\textcolor{black}{For Example \ref{example:555XZ}, the covariant subgroup is the $\mathbb{Z}_5$ qudit Pauli group. Again, the logical representation of the generators is the Pauli $X$ and $Z$ matrices for the $\mathbb{Z}_5$-qudit.
\begin{eqs}
    &\mathcal{P}_{\text{$\mathbb{Z}_5$-qudit}}= \langle X,Z |X^5=Z^5=\Omega X \Omega^{-1}X^{-1}=\Omega Z \Omega^{-1}Z^{-1}=\mathbbm{1}\rangle,\\& \rho_L(Z)=\begin{pmatrix}
        1&0&0&0&0\\0&e^{i\frac{2\pi}{5}}&0&0&0\\0&0&e^{i\frac{4\pi}{5}}&0&0\\0&0&0&e^{i\frac{6\pi}{5}}&0\\0&0&0&0&e^{i\frac{8\pi}{5}}\\
    \end{pmatrix},~\rho_L(X)=\begin{pmatrix}
        0&1&0&0&0\\0&0&1&0&0\\0&0&0&1&0\\0&0&0&0&1\\1&0&0&0&0
    \end{pmatrix}.
\end{eqs}
Similarly, $\Omega\equiv XZX^{-1}Z^{-1}$ is the group commutator of $X$ and $Z$. $\mathcal{P}_{\text{$\mathbb{Z}_5$-qudit}}$ has $125$ elements. In terms of generators, they are written as 
\begin{eqs}\label{eq:ququintelement}
   \mathcal{P}_{\text{$\mathbb{Z}_5$-qudit}}= \{\Omega^i X^j Z^k|\forall i,j,k\in\mathbb{Z}_5\} 
\end{eqs}}

\textcolor{black}{Different from Example \ref{example:333XZ}, the compatible tessellation is $\{5,5,5\}$ on a hyperbolic plane. Its proper triangle group is $\bar{\Delta}(5,5,5)=\langle r_A, r_B|~r_A^5=r_B^5=(r_A r_B)^5=\mathbbm{1}\rangle$. Similarly, the physical representation of the Pauli group generators is $\rho(Z)=r_A, \rho(X)=r_B, \rho((ZX)^{-1})=(r_A r_B)^{-1}$. We need to work out how $r_A,~r_B$ act on hyperbolic planes. We adopt the embedding in (\ref{eq:hyperboilcembedding}). $(r_A r_B)^{-1}$ is the rotation counterclockwise around the central point $(\eta=0)$ by $\frac{2\pi}{5}$, while $r_A$ is the rotation counterclockwise around point $(\eta=b,\theta=0)$ by $\frac{2\pi}{5}$. Here $b$ is determined by the trigonometry on the hyperbolic plane via $\cosh b=\frac{\cos\frac{\pi}{5}+\cos^2\frac{\pi}{5}}{\sin^2\frac{\pi}{5}}$. See (\ref{eq:hyperbolictriangle}). Similarly, the action of $r_A$ is decomposed into first shifting to the origin, then performing the rotation, and finally shifting back to point $A$. If a point on the hyperbolic space is parameterized as $(\cosh{\eta},\sinh{\eta}\cos\theta,\sinh{\eta}\sin\theta)$, then the rotation generators  take the representation as 
\begin{eqs}
     (r_A r_B)^{-1}=\begin{pmatrix}
     1&0&0\\
        0&\cos\frac{2\pi}{5}&-\sin\frac{2\pi}{5}\\ 0&\sin\frac{2\pi}{5}&\cos\frac{2\pi}{5}
    \end{pmatrix},~
    r_A=\begin{pmatrix}
    \cosh b&\sinh b&0\\ \sinh b&\cosh b&0\\0&0&1
    \end{pmatrix}\begin{pmatrix}
     1&0&0\\
        0&\cos\frac{2\pi}{5}&-\sin\frac{2\pi}{5}\\ 0&\sin\frac{2\pi}{5}&\cos\frac{2\pi}{5}
    \end{pmatrix}\begin{pmatrix}
    \cosh b&-\sinh b&0\\ -\sinh b&\cosh b&0\\0&0&1
    \end{pmatrix},
\end{eqs}
and $r_B=((r_A r_B)^{-1}r_A)^{-1}$.}

\textcolor{black}{Similar to the previous example, we need to sum over all equivalent forms of each logical operation. However, on hyperbolic planes, this sum does not take a closed form. Instead, we only present the structure of the unit cell for all examples on hyperbolic planes. In Figure \ref{fig:555QuditXZ} and the following hyperbolic examples, the unit cells are determined with the assistance of the GAP package \textsf{HyperCells} \cite{HyperCells}, the Mathematica package \textsf{HyperBloch} \cite{HyperBloch}, and the data of the quotient of triangle groups in Ref.~\cite{Conder:2007}. Information about the generalized stabilizers can also be calculated using these packages, as long as the logical group structure is identified in Ref.~\cite{Conder:2007}. The group $\mathcal{P}_{\text{$\mathbb{Z}_5$-qudit}}$ is listed as T26.37 in Ref.~\cite{Conder:2007}. To get the codeword configurations in Figure \ref{fig:555QuditXZ}, we choose $\ket{\text{init}}=\ket{(b,0)}$ and enumerate over the group elements to calculate $\bra{0}\rho_L(g)^\dagger\ket{k} \rho(g)\ket{\text{init}}$. We also apply the generalized stabilizer relations to determine the equivalent points in the unit cell.}

\textcolor{black}{\subsubsection{Example \ref{example:468Clifford}}}
\textcolor{black}{The following two examples are technically similar to Example \ref{example:555XZ}. For Example \ref{example:468Clifford}, the covariant subgroup is the qubit Clifford group $\mathcal{C}$. The logical representation of the generators is the $S$ and $U$ matrices.
\begin{eqs}
    \mathcal{C} =\langle S,U|~  S^8=U^6=(US)^{4}=S^4 U^3=S^4 (US)^2=\mathbbm{1}\rangle,~S=\begin{pmatrix}
        e^{i\frac{\pi}{4}}&0\\0&e^{-i\frac{\pi}{4}}\\
    \end{pmatrix},~ U=\frac{1}{\sqrt{2}}\begin{pmatrix}
        e^{i\frac{\pi}{4}}&e^{i\frac{\pi}{4}}\\-e^{-i\frac{\pi}{4}}&e^{-i\frac{\pi}{4}}
    \end{pmatrix}.
\end{eqs}
$\mathcal{C}$ has $48$ elements. In terms of generators, they take the form
\begin{eqs}
   \mathcal{C}= \{S^i, S^i U, S^i U^{-1}, S^i U S^{-1}, S^i U^{-1}S, S^i U S^{-1} U  |\forall i\in\mathbb{Z}_8\} 
\end{eqs}}

\textcolor{black}{The compatible tessellation is $\{6,4,8\}$ on a hyperbolic plane. Its proper triangle group is $\bar{\Delta}(6,4,8)=\langle r_A, r_B|~r_A^6=r_B^4=(r_A r_B)^8=\mathbbm{1}\rangle$. The physical representation of the generators is $\rho(S)=(r_A r_B)^{-1}, \rho(U)=r_A$. With the embedding in (\ref{eq:hyperboilcembedding}), $(r_A r_B)^{-1}$ is the rotation counterclockwise around the central point $(\eta=0)$ by $\frac{\pi}{4}$, while $r_A$ is the rotation counterclockwise around point $(\eta=b,\theta=0)$ by $\frac{\pi}{3}$. $b$ is determined by $\cosh b=\frac{\cos{\frac{\pi}{4}}+\cos \frac{\pi}{8}\cos \frac{\pi}{6}}{\sin \frac{\pi}{8}\sin \frac{\pi}{6}}$. See (\ref{eq:hyperbolictriangle}). The rotation generators take the representation as 
\begin{eqs}
     (r_A r_B)^{-1}=\begin{pmatrix}
     1&0&0\\
        0&\cos\frac{\pi}{4}&-\sin\frac{\pi}{4}\\ 0&\sin\frac{\pi}{4}&\cos\frac{\pi}{4}
    \end{pmatrix},~
    r_A=\begin{pmatrix}
    \cosh b&\sinh b&0\\ \sinh b&\cosh b&0\\0&0&1
    \end{pmatrix}\begin{pmatrix}
     1&0&0\\
        0&\cos\frac{\pi}{3}&-\sin\frac{\pi}{3}\\ 0&\sin\frac{\pi}{3}&\cos\frac{\pi}{3}
    \end{pmatrix}\begin{pmatrix}
    \cosh b&-\sinh b&0\\ -\sinh b&\cosh b&0\\0&0&1
    \end{pmatrix},
\end{eqs}
and $r_B=((r_A r_B)^{-1}r_A)^{-1}$.}

\textcolor{black}{The group $\mathcal{C}$ is listed as T12.21 in Ref.~\cite{Conder:2007}. To get the codeword configurations in Figure \ref{fig:468Cliffordhyperbolic}, we choose $\ket{\text{init}}=\ket{(\frac b2,0)}$ and enumerate over the group elements to calculate $\bra{0}\rho_L(g)^\dagger\ket{k} \rho(g)\ket{\text{init}}$ for each logical state $\ket{k}$, and superpose them correspondingly to get the codeword. }

\textcolor{black}{\subsubsection{Example \ref{example:4352I}}}
For Example \ref{example:4352I}, the covariant subgroup is the binary icosahedral group  $\mathcal{I}_2$. We write their generators as $g_1,~g_2$ for simplicity. In concordance with the literature, they are related to the $F$ and $\Phi$ matrices as follows. 
\begin{eqs}
&\mathcal{I}_2 =\langle g_1, g_2|~  g_1^4=g_2^3=(g_1 g_2)^5=g_1^2 g_2 g_1^2 g_2^2=\mathbbm{1}\rangle, \rho_L(g_1)=\Phi F^{-1},~\rho_L(g_2)=-F, \rho_L(g_1 g_2)=-\Phi,
\\&F=\frac{e^{-i\frac{\pi}{4}}}{\sqrt{2}}\begin{pmatrix}
        1&-i\\1&i\\
    \end{pmatrix},~ \Phi=\frac{1}{2}\begin{pmatrix}
        \phi_G+ i\phi_G^{-1}&1\\-1&\phi_G- i\phi_G^{-1}
    \end{pmatrix}.
\end{eqs}
Here $\phi_G=\frac{\sqrt{5}+1}{2}$ is the golden ratio.
$\mathcal{I}_2$ has $120$ elements.

\textcolor{black}{The compatible tessellation is $\{4,3,5\}$ on a hyperbolic plane. Its proper triangle group is $\bar{\Delta}(4,3,5)=\langle r_A, r_B|~r_A^4=r_B^3=(r_A r_B)^5=\mathbbm{1}\rangle$. The physical representation of the generators is $\rho(-\Phi^{-1})=(r_A r_B)^{-1},~\rho(\Phi F^{-1})=r_A,~\rho(-F)=r_B$. With the embedding in (\ref{eq:hyperboilcembedding}), $(r_A r_B)^{-1}$ is the rotation counterclockwise around the central point $(\eta=0)$ by $\frac{2\pi}{5}$, while $r_A$ is the rotation counterclockwise around point $(\eta=b,\theta=0)$ by $\frac{\pi}{2}$. $b$ is determined by $\cosh b=\frac{\cos{\frac{\pi}{3}}+\cos \frac{\pi}{4}\cos \frac{\pi}{5}}{\sin \frac{\pi}{4}\sin \frac{\pi}{5}}$. The rotation generators take the representation as 
\begin{eqs}
     (r_A r_B)^{-1}=\begin{pmatrix}
     1&0&0\\
        0&\cos\frac{2\pi}{5}&-\sin\frac{2\pi}{5}\\ 0&\sin\frac{2\pi}{5}&\cos\frac{2\pi}{5}
    \end{pmatrix},~
    r_A=\begin{pmatrix}
    \cosh b&\sinh b&0\\ \sinh b&\cosh b&0\\0&0&1
    \end{pmatrix}\begin{pmatrix}
     1&0&0\\
        0&\cos\frac{\pi}{2}&-\sin\frac{\pi}{2}\\ 0&\sin\frac{\pi}{2}&\cos\frac{\pi}{2}
    \end{pmatrix}\begin{pmatrix}
    \cosh b&-\sinh b&0\\ -\sinh b&\cosh b&0\\0&0&1
    \end{pmatrix},
\end{eqs}
and $r_B=((r_A r_B)^{-1}r_A)^{-1}$.}

\textcolor{black}{The group $\mathcal{I}_2$ is listed as T14.14 in Ref.~\cite{Conder:2007}. To get the codeword configurations in Figure \ref{fig:468Cliffordhyperbolic}, we choose $\ket{\text{init}}=\ket{(\frac b2,0)}$ and enumerate over the group elements to calculate $\bra{0}\rho_L(g)^\dagger \ket{k} \rho(g)\ket{\text{init}}$ for each logical state $\ket{k}$, and superpose them correspondingly to obtain the codeword.}

\section{Analysis of momentum error }\label{app:momentum}
We denote the position basis of the two-dimensional surfaces as $\ket{x}$, which is normalized to $\langle x|x^\prime\rangle=\frac{\delta(x_1-x_1^\prime)\delta(x_2-x_2^\prime)}{\sqrt{h(x)}}\equiv \delta^{(2)}(x-x^\prime)$. Here $h(x)$ is the determinant of the metric evaluated at the point $x$. In particular
\begin{itemize}
    \item For the sphere:  $\langle \theta,\phi|\theta^\prime,\phi^\prime\rangle=\frac{\delta(\theta-\theta^\prime)\delta(\phi-\phi^\prime)}{\sin\theta}$.
    \item For the Euclidean plane: $\langle x,y|x^\prime,x^\prime\rangle=\delta(x-x^\prime)\delta(y-y^\prime)$. 
    \item For the hyperbolic plane: $\langle \eta,\theta|\eta^\prime,\theta^\prime\rangle=\frac{\delta(\theta-\theta^\prime)\delta(\eta-\eta^\prime)}{\sinh \eta}$.
\end{itemize}
Under any isometry $\rho(g)$ of the surface, the delta function is invariant: $\delta^{(2)}(x-x^\prime)=\delta^{(2)}(\rho(g)x-\rho(g)x^\prime)$. 

In the position basis, an error operator can be written as
\begin{equation}\label{eq:momentumerror}
    \hat{E}_{r,n}=\int d^2x \sqrt{h(x)}  \langle x \ket{r,n} \ket{x}\bra{x},
\end{equation}
where $r$ labels the representation and $n$ labels the basis within the representation. This notation matches the momentum errors in Table~\ref{tab:errortype}:
\begin{itemize}
    \item Sphere: $r\equiv \ell, n\equiv m$ in $Y_\ell^m$.
    \item Euclidean plane: $r\equiv \{k_x, k_y\}$ in $V_{\vec{k}}$, with one-dimensional representations (no $n$).
    \item Hyperbolic plane: $r \equiv s$ and $n$ corresponds to the $n$ in $P_{-\frac 12+i r}^n(\cosh \hat{\rho})e^{i n \hat\phi}$. 
\end{itemize}
The error correction condition is
\begin{eqs}\label{eq:momentumKL}
\bra{\bar{i}}\hat{E}^{\dagger}_{r_1,n_1}\hat{E}_{r_2,n_2}\ket{\bar{j}}= \int d^2x  \sqrt{h(x)} \langle x \ket{r_2,n_2}\bra{r_1,n_1} x\rangle \langle \bar{i}\ket{x}\bra{x} \bar j\rangle. 
\end{eqs}
Following Eq.~\eqref{eq:encodingmap}, we obtain 
\begin{eqs}
    \bra{x}\bar{j}\rangle&=\sum_{\gamma\in\Gamma}\sum_{g\in G} 
    \bra{\Sigma} \rho^\dagger_L(g)\ket{j}\bra{x}\rho(\gamma g_0)\ket{p_i}=\sum_{\gamma\in\Gamma}\sum_{g\in G} 
    \bra{\Sigma} \rho^\dagger_L(g)\ket{i}\delta^{(2)}(x-\rho(\gamma g_0)p_i),\\
    \bra{x}\bar{j}\rangle\bra{\bar{i}}x\rangle&=\sum_{\gamma_1, \gamma_2\in\Gamma}\sum_{g_1,g_2\in G} \bra{\Sigma} \rho^\dagger_L(g_1)\ket{j}\bra{i} \rho_L(g_2)\ket{\Sigma}\delta^{(2)}(x-\rho(\gamma_1 g_{01})p_i)\delta^{(2)}(x-\rho(\gamma_2 g_{02})p_i)\\
    &=\sum_{\gamma\in\Gamma}\sum_{g_1,g_2\in G} \bra{\Sigma} \rho^\dagger_L(g_1)\ket{j}\bra{i} \rho_L(g_2)\ket{\Sigma} \delta^{(2)}(x-\rho(\gamma_1 g_{01})p_i)\delta^{(2)}(\rho(\gamma g_{01})p_i-\rho(\gamma g_{02})p_i)\\
    &=\sum_{\gamma\in\Gamma}\sum_{g_1,g_2\in G} \bra{\Sigma} \rho^\dagger_L(g_1)\ket{j}\bra{i} \rho_L(g_2)\ket{\Sigma} \delta^{(2)}(x-\rho(\gamma g_{01})p_i)\delta^{(2)}(p_i-\rho(g_{01}^{-1} g_{02})p_i)\\
    &=\sum_{\gamma\in\Gamma}\sum_{g_1\in G}\sum_{g_{p}\in F_{p_i}} \bra{\Sigma} \rho^\dagger_L(g_1)\ket{j}\bra{i} \rho_L(g_1)\rho_L(g_p)\ket{\Sigma} \delta^{(2)}(x-\rho(\gamma g_{01})p_i)\delta^{(2)}(p_i-p_i).\\
\end{eqs}
Here, in the second line, the sums over $\gamma_1$,$\gamma_2$ reduce to one because different unit cells do not overlap. The third line follows from the covariance of the $\delta^{(2)}$ function under an isometry of the surface. The fourth line follows from a change of summing variable from $g_2$ to $g_p$. The sum over $g_p$ further reduces from $G$ to a subgroup which keeps $p_i$ invariant, denoted as $F_{p_i}$. 

If $F_{p_i}$ is nontrivial, since $\rho(g_p)p_i=p_i$, the delta function $\delta^{(2)}(p_i-\rho(g_p)p_i)$ becomes $\delta^{(2)}(p_i-p_i)$. The sum $\sum_{g_p\in F_{p_i}}\rho_L(g_p)$ is proportional to the projector onto the common eigenvalue-$1$ subspace of all the group elements $\rho_L(g_p)$. As mentioned in the main article, in this case, we choose $\ket{\Sigma}$ to be in this subspace and 
\begin{eqs}
\bra{x}\bar{j}\rangle\bra{\bar{i}}x\rangle=|F_{p_i}|\delta^{(2)}(p_i-p_i)\sum_{\gamma\in\Gamma}\sum_{g_1\in G}\bra{\Sigma} \rho^\dagger_L(g_1)\ket{j}\bra{i} \rho_L(g_1)\ket{\Sigma} \delta^{(2)}(x-\rho(\gamma_1 g_{01})p_i). 
\end{eqs}
If $F_{p_i}$ is trivial, then $|F_{p_i}|=1$ and the above equation also applies, but for an arbitrary $\ket{\Sigma}$.

Substituting back into Eq.~\eqref{eq:momentumKL}, we obtain 
\begin{eqs} \label{eq:momentumKL2}
\bra{\bar{i}}\hat{E}^{\dagger}_{r_1,n_1}\hat{E}_{r_2,n_2}\ket{\bar{j}}&=|F_{p_i}|\delta^{(2)}(p_i-p_i)\sum_{\gamma\in\Gamma}\sum_{g\in G}\bra{\Sigma} \rho^\dagger_L(g)\ket{j}\bra{i} \rho_L(g)\ket{\Sigma}\langle \rho(\gamma g_{0})p_i \ket{r_2,n_2}\bra{r_1,n_1} \rho(\gamma g_{0})p_i\rangle. 
\end{eqs}

If $\Gamma$ is trivial, no further simplification is possible, and Eq.~\eqref{eq:momentumKL2} must be evaluated using the explicit data of the specific code. This is the case for Example \ref{example:224XZ}, with the detailed calculation provided in Section \ref{sec:224XZmomentum}. 

If $\Gamma$ is nontrivial, it forms a discrete subgroup of the full isometry group of the surface. Recall that $\ket{r_1,n_1},~\ket{r_2,n_2}$ are the bases of the unitary irreducible representations of the full isometry group. While representations $r_1,~r_2$ are also unitary representations for the subgroup $\Gamma$, they may not remain irreducible. In the cases of our interests, when $\Gamma$ is the translation group on the Euclidean or hyperbolic surface, the representations $r_1,~r_2$ remain irreducible. Therefore, we can first define
\begin{eqs}
V_{\Gamma}\equiv\sum_{\gamma\in\Gamma}\rho(\gamma)^{-1}\ket{r_2,n_2}\bra{r_1,n_1}\rho(\gamma),    
\end{eqs}
and apply Schur's lemma. Three scenarios arise:
\begin{enumerate}
    \item \textbf{Identical representations and bases:} If $r_1=r_2$ and $n_1=n_2$, then $V_\Gamma=c(r_1,n_1)\mathbb{1}$. Eq. \eqref{eq:momentumKL2} becomes
\begin{eqs}
    \bra{\bar{i}}\hat{E}^{\dagger}_{r_1,n_1}\hat{E}_{r_1,n_1}\ket{\bar{j}}&=|F_{p_i}|\delta^{(2)}(p_i-p_i)c(r_1,n_1)\sum_{g\in G}\bra{\Sigma} \rho^\dagger_L(g)\ket{j}\bra{i} \rho_L(g)\ket{\Sigma} \delta^{(2)}(\rho(g_0)p_i-\rho(g_0)p_i)\\
    &=|F_{p_i}||G|\left(\delta^{(2)}(p_i-p_i)\right)^2c(r_1,n_1) \delta_{i,j},
\end{eqs}
where the second equality uses the invariance of the $\delta^{(2)}$ under isometries and Schur's lemma for the logical group $G$. Therefore, the KL condition is satisfied.

    \item \textbf{Inequivalent representations or bases:} If $r_1,~r_2$ are the nonequivalent representations or $n_1,~n_2$ are nonequivalent bases, Schur’s lemma guarantees
    \begin{eqs}
        \bra{\bar{i}}\hat{E}^{\dagger}_{r_1,n_1}\hat{E}_{r_2,n_2}\ket{\bar{j}}=0,\quad \forall i,j,
    \end{eqs}
    so the error correction condition is also satisfied.

    \item \textbf{Potential violation:} If $\ket{r_1,n_1},~\ket{r_2,n_2}$ are different bases for nonequivalent representations for the full isometry group but become equivalent when restricted to the subgroup $\Gamma$, the error correction condition may fail. In this case, Eq.~\eqref{eq:momentumKL2} can be written as
    \begin{eqs}
\bra{\bar{i}}\hat{E}^{\dagger}_{r_1,n_1}\hat{E}_{r_1,n_1}\ket{\bar{j}}&=|F_{p_i}|\left(\delta^{(2)}(p_i-p_i)\right)^2\sum_{g\in G}c(r_1,n_1,r_2,n_2, g_0 p_i)\bra{\Sigma} \rho^\dagger_L(g)\ket{j}\bra{i} \rho_L(g)\ket{\Sigma}.
\end{eqs}
In this scenario, one must plug in the specific code data to determine whether the error correction condition is violated for the given choice of the pair of errors.
\end{enumerate}

\subsection{On correction of momentum error}
We now discuss active error correction for momentum errors. The approach is analogous to that used in GKP codes: express the codeword in the momentum basis and examine how it transforms under the action of an error operator. As we shall see, the more complicated structure of representation theory for the isometry group on curved spaces makes error correction for momentum errors less straightforward.

Consider a momentum error $\hat{E}_{r_0,n_0}$ acts on the codeword $\ket{\bar i}$, we have
\begin{eqs}\label{eq:momentumerroroncodeword}
    \hat{E}_{r_0,n_0}\ket{\bar i}&\propto \int d^2x \sqrt{h(x)}  \langle x \ket{r_0,n_0} \ket{x}\bra{x}\bar{i}\rangle=\sum_{r,n} \int d^2x \sqrt{h(x)}  \langle x \ket{r_0,n_0} \bra{x}r,n\rangle\bra{r,n}\bar{i}\rangle\ket{x}\\
    &=\sum_{r,n} \int d^2x \sqrt{h(x)} \sum_{r^\prime, n^\prime} c_{r_0,n_0;r,n}^{r^\prime, n^\prime} \langle x \ket{r^\prime,n^\prime}\bra{r,n}\bar{i}\rangle\ket{x}=\sum_{r,n}\sum_{r^\prime, n^\prime}  c_{r_0,n_0;r,n}^{r^\prime, n^\prime} \ket{r^\prime,n^\prime}\bra{r,n}\bar{i}\rangle .
\end{eqs}
In the second line, we perform the decomposition
\begin{equation} \label{eq:momentumerrordecompose}
    \langle x \ket{r_0,n_0} \bra{x}r,n\rangle=\sum_{r^\prime, n^\prime} c_{r_0,n_0;r,n}^{r^\prime, n^\prime} \langle x \ket{r^\prime,n^\prime}. 
\end{equation}
This is the standard problem of decomposing the tensor product of two irreducible representations into a direct sum of irreducible representations. 
For the spherical case whose isometry group is $\text{SO}(3)$, the irreducible representation labels $(r,n)$ correspond to angular and magnetic quantum numbers $(\ell,m)$. This reduces to the standard Clebsch--Gordan problem of combining two angular momenta in $\text{SO}(3)$, with the Clebsch--Gordan coefficients $c_{r_0,n_0; r,n}^{r',n'}$ specifying the decomposition. The Clebsch--Gordan coefficients are non-zero only when $|r-r_0| \leq r' \leq r + r_0$ and $n' = n + n_0$. For the hyperbolic case, the isometry group is SL(2,$\mathbb{R}$), and a similar decomposition problem has also been studied; see e.g.,~Ref.~\cite{repka1978tensor}. In general, the Clebsch--Gordan problem for decomposing irreducible representations of general groups becomes more challenging.

The complicated structure of the representation theory of non-Abelian isometry groups for curved space induces challenges to momentum error correction. The challenges primarily stem from two aspects. 

First, syndrome extraction becomes more difficult. The momentum errors are the basis of multi-dimensional irreducible representations, and carry both representation and basis labels $(r,n)$. Using the $\mathrm{C}Z$ logical gate introduced in the main text, we can extract the $n$-syndrome of the error. However, how to extract $r$ remains unclear. It corresponds to the total angular momentum in the spherical case and is related to the total energy in the hyperbolic case. Moreover, non-commutative generalized stabilizers might make the construction of the syndrome extraction circuit more challenging. 

Second, even if the most probable error is identified, the error correction procedure becomes more complicated. When we do error correction, we apply another momentum operator $\hat{E}_{r_c,n_c}$ and the corrected state is $\hat{E}_{r_c,n_c}\hat{E}_{r_0,n_0}\ket{\bar k}$. We need to perform the decomposition similar to that in Eq.~(\ref{eq:momentumerrordecompose}). A successful error correction requires $c_{r_0,n_0,r_c,n_c}^{0,0}\neq 0$, ensuring that the decomposition contains a component of the trivial representation and is thus proportional to the original codeword. However, in general cases, the coefficients $c_{r_0,n_0,r_c,n_c}^{r',n'}$ are not necessarily delta functions, meaning that the error correction only probabilistically recovers the original codeword. Therefore, the error correction procedure for general tessellation codes is inherently more challenging.

For the planar case, both Eqs.~\eqref{eq:momentumerroroncodeword} and \eqref{eq:momentumerrordecompose} can be greatly simplified. This is the situation of the momentum error for Example~\ref{example:333XZ} in our work and the well-studied Gottesman--Kitaev--Preskill (GKP) codes \cite{gkp}. Each irreducible representation is one-dimensional, labeled by $\vec{r}=(k_x,k_y)$ with $ \forall k_x,k_y\in \mathbb{R}$, without extra $n$ label, and satisfies $\langle x \ket{\vec{r}}=e^{i\vec x\cdot\vec r}$. The decomposition of irreducible representations reduces to addition of momentum vectors: 
\begin{eqs}
   \langle x \ket{r_0} \bra{x}r\rangle=\langle x \ket{\vec{r}+\vec{r_0}},\quad \hat{E}_{\vec{r_0}}\ket{\bar i}=\sum_{\vec r}\ket{\vec{r_0}+\vec{r}}\bra{\vec{r}}\bar i\rangle,
\end{eqs}
where $c_{\vec{r}_0,\vec{r}}^{\vec{r}'}=\delta_{\vec{r}', \vec{r}_0+\vec{r}}$. So a momentum error $\hat{E}_{\vec{r}_0}$ shifts momentum eigenstate $\ket{\vec{r}}$ to $\ket{\vec{r}+\vec{r_0}}$. The GKP codewords are the superposition of momentum eigenstates with evenly spaced eigenvalues, so the error correction can be implemented by performing modular measurements in the momentum basis followed by feedback.  Also, the simple structure of the momentum addition rule guarantees that the final state after error correction has only the component of the original codeword.
These properties make momentum error correction straightforward for codes defined on Euclidean planes.

\subsection{Momentum error analysis of Example \ref{example:224XZ}}\label{sec:224XZmomentum}
For a generic codewords in Eq. \eqref{eq:224_codewords}, we can calculate
\begin{eqs}\label{eq:sphericalmomentumerror}
    \bra{\bar 0} Y_{l_2}^{m_2 \dagger} Y_{l_1}^{m_1} \ket{\bar 0}&= \frac{1}{4} \Big( P_{l_1}^{m_1}(\cos\theta_0)P_{l_2}^{m_2}(\cos\theta_0) e^{i(m_2-m_1)\phi_0} + P_{l_1}^{m_1}(\cos\theta_0)P_{l_2}^{m_2}(\cos\theta_0) e^{i(m_2-m_1)(\phi_0+\pi)}\\
    &+P_{l_1}^{m_1}(\cos(\pi-\theta_0))P_{l_2}^{m_2}(\cos(\pi-\theta_0)) e^{-i(m_2-m_1)\phi_0}+P_{l_1}^{m_1}(\cos(\pi-\theta_0))P_{l_2}^{m_2}(\cos(\pi-\theta_0)) e^{i(m_2-m_1)(\pi-\phi_0)}\Big)\\
    &=\frac{1}{4}\Big(P_{l_1}^{m_1}(\cos\theta_0)P_{l_2}^{m_2}(\cos\theta_0) (1+(-1)^{m_1+m_2})(e^{i(m_2-m_1)\phi_0}+e^{-i(m_2-m_1)\phi_0}(-1)^{l_1+l_2})\Big),\\
    \bra{\bar 1} Y_{l_2}^{m_2 \dagger} Y_{l_1}^{m_1} \ket{\bar 1} &=\frac{1}{4}\Big(P_{l_1}^{m_1}(\cos\theta_0)P_{l_2}^{m_2}(\cos\theta_0) (1+(-1)^{m_1+m_2})(e^{i(m_2-m_1)(\phi_0+\frac{\pi}{2})}+e^{-i(m_2-m_1)(\phi_0+\frac{\pi}{2})}(-1)^{l_1+l_2})\Big).
\end{eqs}
We used the property of the associated Legendre polynomials $P_l^m (x)=(-1)^{l+m} P_l^m (-x)$ to get the final expression. The expression for $\ket{\bar 1}$ is different from that of $\ket{\bar 1}$ by $\phi_0\to\phi_0+\frac{\pi}{2}$. We immediately see because of the factor $(1+(-1)^{m_1+m_2})$ and the phase difference $\frac{\pi}{2}$, $\bra{\bar 0} Y_{l_2}^{m_2 \dagger} Y_{l_1}^{m_1} \ket{\bar 0}\neq \bra{\bar 1} Y_{l_2}^{m_2 \dagger} Y_{l_1}^{m_1} \ket{\bar 1}$ may only occur when $(m_2-m_1)\mod 4=2$. In this case, we have 
\begin{eqs}\label{eq:224_detection_condition}
    \langle \overline{0} | Y_{l_2}^{m_2 \dagger} Y_{l_1}^{m_1} | \overline{0} \rangle=-\langle \overline{1} | Y_{l_2}^{m_2 \dagger} Y_{l_1}^{m_1} | \overline{1} \rangle
    =\frac{1}{2}\Big(P_{l_1}^{m_1}(\cos\theta_0)P_{l_2}^{m_2}(\cos\theta_0) (e^{i(m_2-m_1)\phi_0}+e^{-i(m_2-m_1)\phi_0}(-1)^{l_1+l_2})\Big),\\
\end{eqs}
In the case we illustrated in Figure \ref{fig:224XZsphere}, we take $\theta_0=\arccos \frac{1}{\sqrt{3}}$ and $\phi_0=\frac{\pi}{4}$. In this case, we get $ \langle \overline{0} | Y_{l_2}^{m_2 \dagger} Y_{l_1}^{m_1} | \overline{0}\propto (1-(-1)^{l_1+l_2})$. Therefore, it is only non-zero when $l_1+l_2$ is odd. In summary, the KL condition is only violated when $l_1+l_2$ is odd and $(m_2-m_1)\mod 4=2$. Hence, $Y_1^0 Y_2^{\pm 2}$ is the undetectable error with the smallest $|l_1|+|l_2|+|m_1|+|m_2|=5$ .

\subsection{Momentum error analysis of Example \ref{example:244XZ}}\label{sec:244XZmomentum}

For the code of Example \ref{example:244XZ}, we calculate the KL condition for a generic code state in Eq. \eqref{eq:Euclideantessellationcodeword} as
\begin{eqs}
    &\langle \overline{0}|\hat{V}_{\Vec{k}_2}^\dagger \hat{V}_{\Vec{k}_1}| \overline{0}\rangle\\
    =&\sum_{n_x, n_y\in \mathbb{Z}} e^{4i (n_x \Delta k_x+ n_y \Delta k_y)} \Big( 4 \cos{(\Delta k_x+\Delta k_y)} \big(\cos{(\Delta k_x -\Delta k_y + x_0 \Delta k_y -y_0\Delta k_x)}+\cos{(\Delta k_x +\Delta k_y - x_0 \Delta k_x -y_0\Delta k_y})\big) \Big)\\
    =& \delta_{\pi/2}( \Delta k_x ) \delta_{\pi/2}( \Delta k_y) \Big( 4 \cos{(\Delta k_x+\Delta k_y)} \big(\cos{(\Delta k_x -\Delta k_y + x_0 \Delta k_y -y_0\Delta k_x)}+\cos{(\Delta k_x +\Delta k_y - x_0 \Delta k_x -y_0\Delta k_y})\big) \Big),\\
    &\cos{(\Delta k_x+\Delta k_y)}\langle\overline{1}|\hat{V}_{\Vec{k}_2}^\dagger \hat{V}_{\Vec{k}_1}| \overline{1}\rangle= \cos{(\Delta k_x-\Delta k_y)}\langle \overline{0}|\hat{V}_{\Vec{k}_2}^\dagger \hat{V}_{\Vec{k}_1}| \overline{0}\rangle,
\end{eqs}
where $\Delta k_x= k_{x1}-k_{x2}, \Delta k_y= k_{y1}-k_{y2}, \delta_{\pi/2}(x)=\sum_{l \in \mathbb{Z}} \delta(x-\frac{l \pi}{2})$. Because of the delta functions, these two values are non-zero only if $\Delta k_x, \Delta k_y$ are multiplicities of $\frac{\pi}{2}$. Further evaluating the expression when $\Delta k_x, \Delta k_y$ are multiplicities of $\frac{\pi}{2}$, we see that the KL condition is violated only when $\Delta k_x, \Delta k_y$ are both odd multiplicities of $\frac{\pi}{2}$. These circumstances correspond to the logical Pauli $Z$ operator.

\subsection{Momentum error analysis of Example \ref{example:333XZ}}\label{sec:333XZmomentum}
For the codewords of Example \ref{example:333XZ} in Eq.~\eqref{eq:honeycombcodewords}, the KL condition is written as
\begin{eqs}
     \langle \overline{0}|\hat{V}_{\Vec{k}_2}^\dagger \hat{V}_{\Vec{k}_1}| \overline{0}\rangle
     &= (e^{i\Delta k_x}+ e^{-2i\Delta k_x} + e^{i (-\frac{1}{2}\Delta k_x + \frac{3\sqrt{3}}{2} \Delta k_y)}) \sum_{m,n \in \mathbb{Z}} e^{i [(\frac{9}{2} \Delta k_x +\frac{3\sqrt{3}}{2} \Delta k_y)m+ 3\sqrt{3} n\Delta k_y]},\\
     &= (e^{i\Delta k_x}+ e^{-2i\Delta k_x} + e^{i (-\frac{1}{2}\Delta k_x + \frac{3\sqrt{3}}{2} \Delta k_y)}) \delta_{2\pi}(\frac{9}{2} \Delta k_x +\frac{3\sqrt{3}}{2} \Delta k_y)\delta_{2\pi}(3\sqrt{3}\Delta k_y),\\
     \langle \overline{1}|\hat{V}_{\Vec{k}_2}^\dagger \hat{V}_{\Vec{k}_1}| \overline{1}\rangle 
     &= e^{i (\frac{3}{2}\Delta k_x-\frac{\sqrt{3}}{2}\Delta k_y)} \langle \overline{0}|\hat{V}_{\Vec{k}_2}^\dagger \hat{V}_{\Vec{k}_1}| \overline{0}\rangle,\\
     \langle \overline{2}|\hat{V}_{\Vec{k}_2}^\dagger \hat{V}_{\Vec{k}_1}| \overline{2}\rangle &= e^{i (-\sqrt{3}\Delta k_y)} \langle \overline{0}|\hat{V}_{\Vec{k}_2}^\dagger \hat{V}_{\Vec{k}_1}| \overline{0}\rangle.\\
\end{eqs}

Because of the periodic delta function, $\bra{\bar{0}} \hat{V}_{\Vec{k}_2}^\dagger \hat{V}_{\Vec{k}_1} \ket{\bar{0}}\neq 0$ only when $\frac{9}{2} \Delta k_x +\frac{3\sqrt{3}}{2} \Delta k_y=2m\pi$ and $3\sqrt{3}\Delta k_y=2n\pi$, $m,n\in \mathbb{Z}$. In these cases $\bra{\bar{1}} \hat{V}_{\Vec{k}_2}^\dagger \hat{V}_{\Vec{k}_1}\ket{\bar{1}}=e^{i\frac{2\pi}{3}(m-n)}\bra{\bar{0}} \hat{V}_{\Vec{k}_2}^\dagger \hat{V}_{\Vec{k}_1} \ket{\bar{0}}$ and $\bra{\bar{2}} \hat{V}_{\Vec{k}_2}^\dagger \hat{V}_{\Vec{k}_1}\ket{\bar{2}}=e^{-i\frac{2\pi}{3}n}\bra{\bar{0}} \hat{V}_{\Vec{k}_2}^\dagger \hat{V}_{\Vec{k}_1}\ket{\bar{0}}$. Therefore, the KL condition is violated when either $m$ or $n$ is not a multiple of $3$. They are equivalent to the cases when both $3\Delta k_x$ and $\sqrt{3} \Delta k_y$ are multiplicities of $\frac{2\pi}{3}$ but not multiplicities of $2\pi$, as stated in the main text.

\section{Figures of code state constellations}
In this section, we collect the figures (Figures \ref{fig:555QuditXZ}, \ref{fig:468Cliffordhyperbolic} and \ref{fig:4352Ihyperbolic}) which illustrate the code state constellation configurations in Examples \ref{example:555XZ},~\ref{example:468Clifford} and \ref{example:4352I}, respectively.
\begin{figure}
\centering
\includegraphics[width=0.75\textwidth]{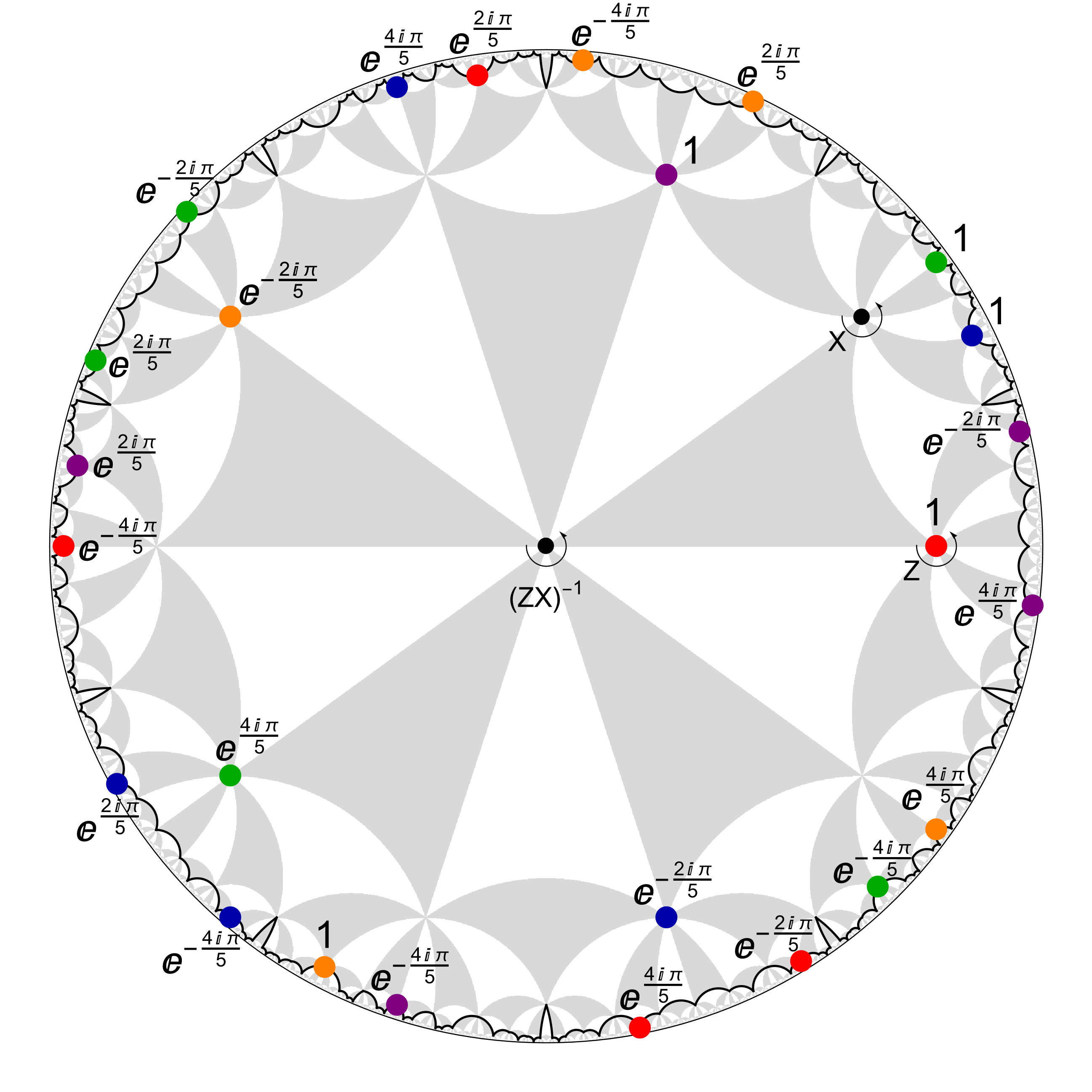}
    \caption{The constellation of the hyperbolic code in Example \ref{example:555XZ}, designed to realize $\mathbb{Z}_5$ qudit Pauli group via rotation. The red, blue, green, orange and purple colours label the logical $\ket{\bar{0}}$ to $\ket{\bar{4}}$ respectively. The coefficients in the superposition are labelled next to the point.}
    \label{fig:555QuditXZ}
\end{figure}

\begin{figure}[tbh]
\centering
\includegraphics[width=0.65\textwidth]{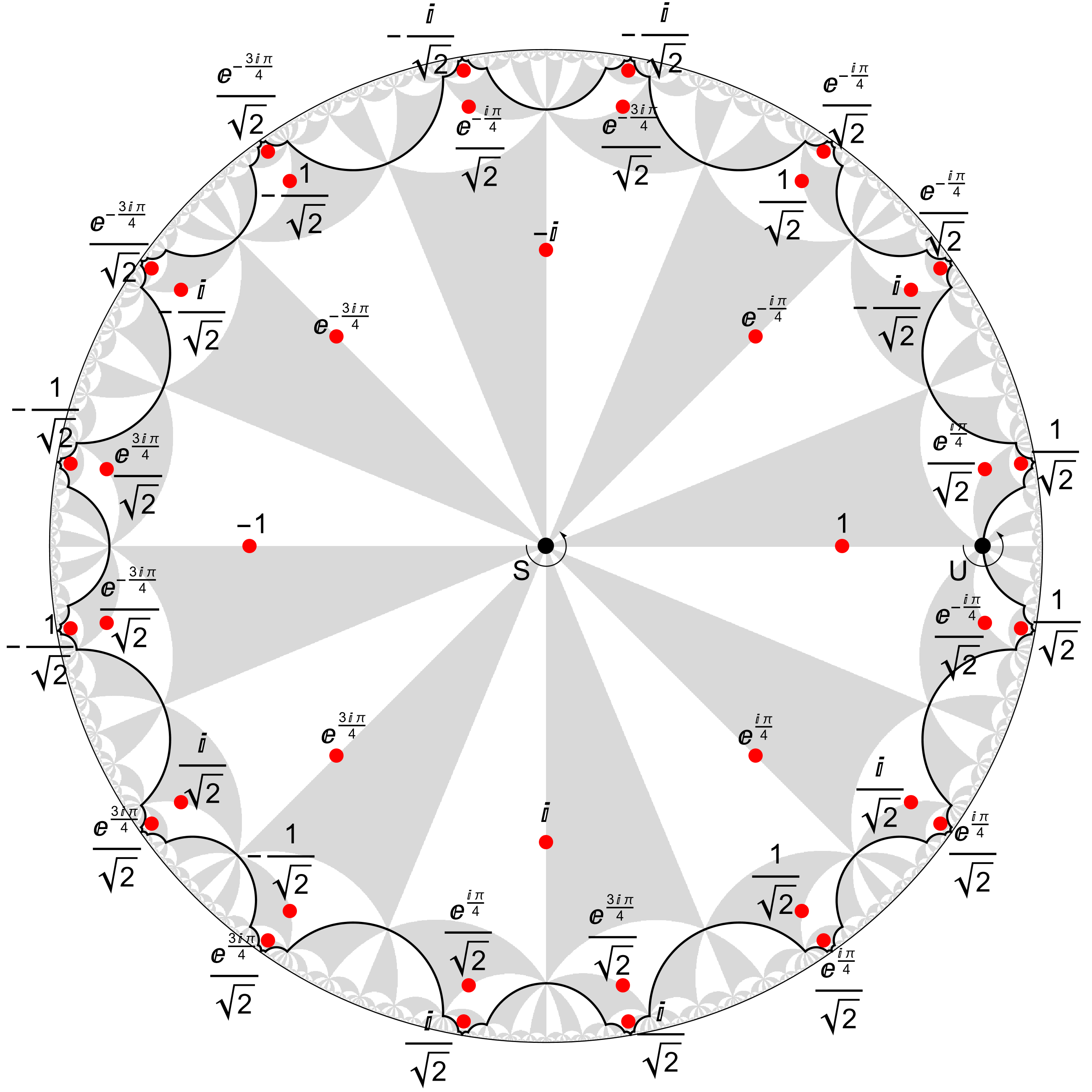}
\includegraphics[width=0.65\textwidth]{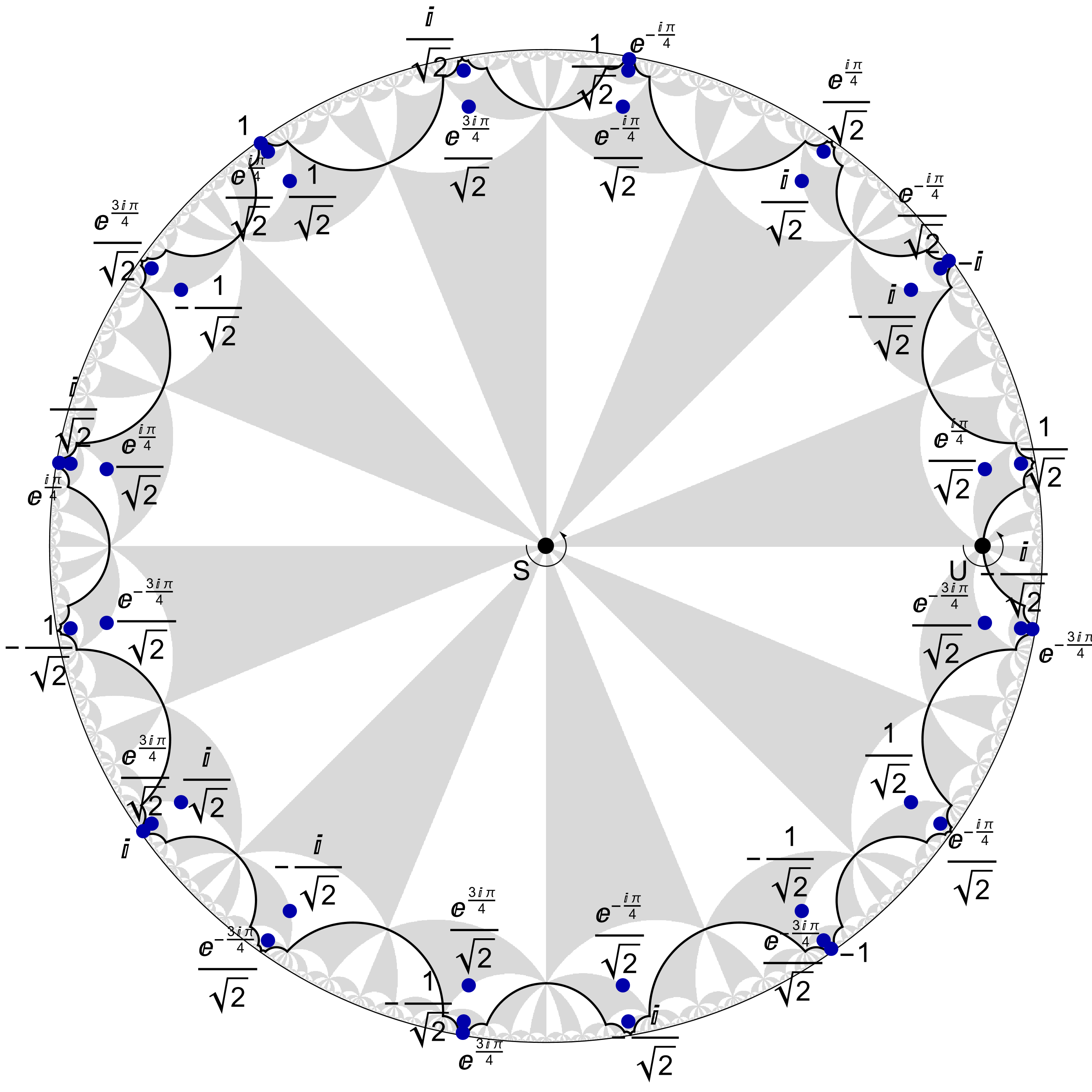}
    \caption{The upper and lower graphs illustrate the codeword configuration of the logical $\ket{\bar{0}}$ and $\ket{\bar{1}}$ states for the construction in Example \ref{example:468Clifford} respectively. Each codeword has a superposition of $40$ points in one unit cell. The coefficients of the superposition are labelled next to the points.}
    \label{fig:468Cliffordhyperbolic}
\end{figure}

\begin{figure}[tbh]
\centering
\includegraphics[width=0.65\textwidth]{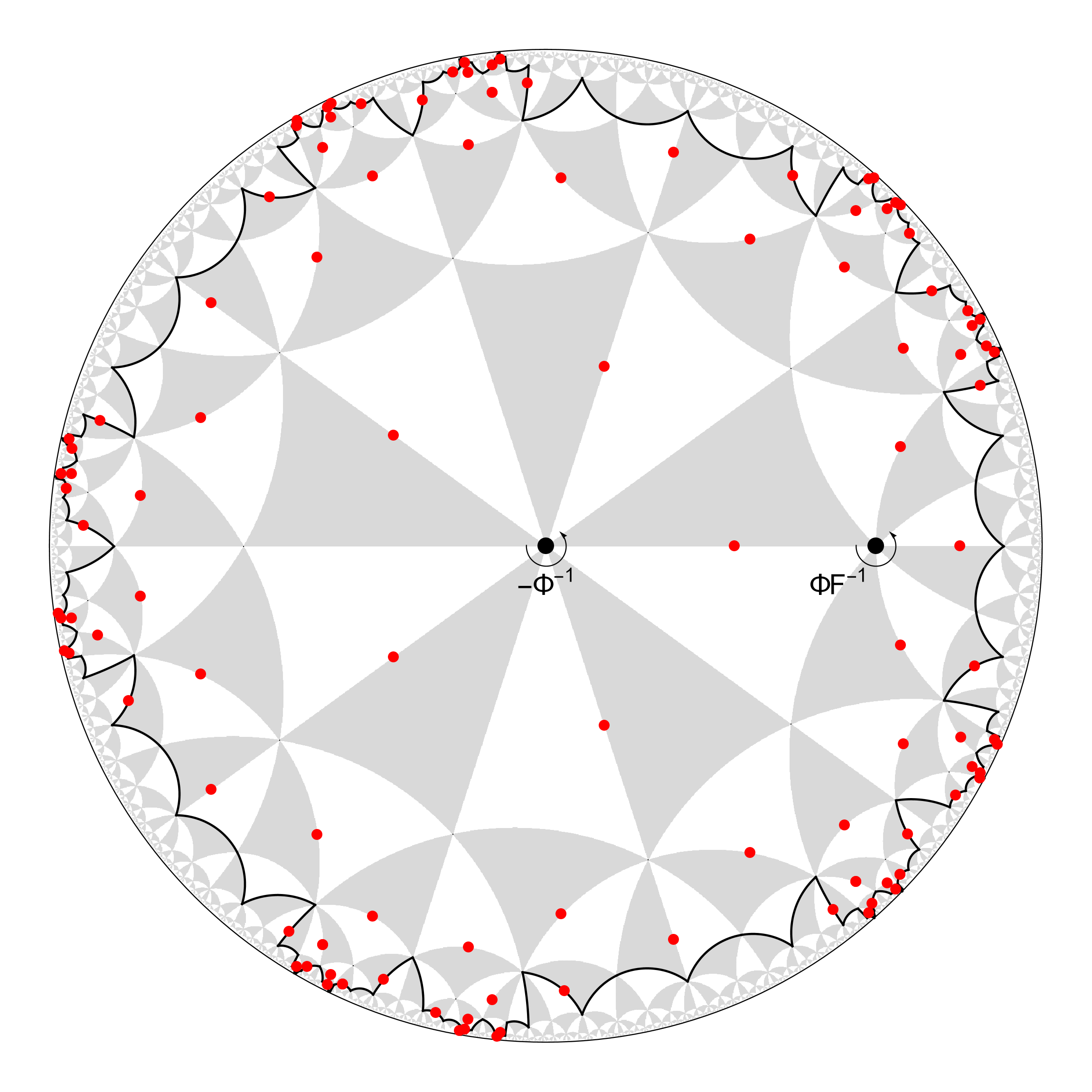}
\includegraphics[width=0.65\textwidth]{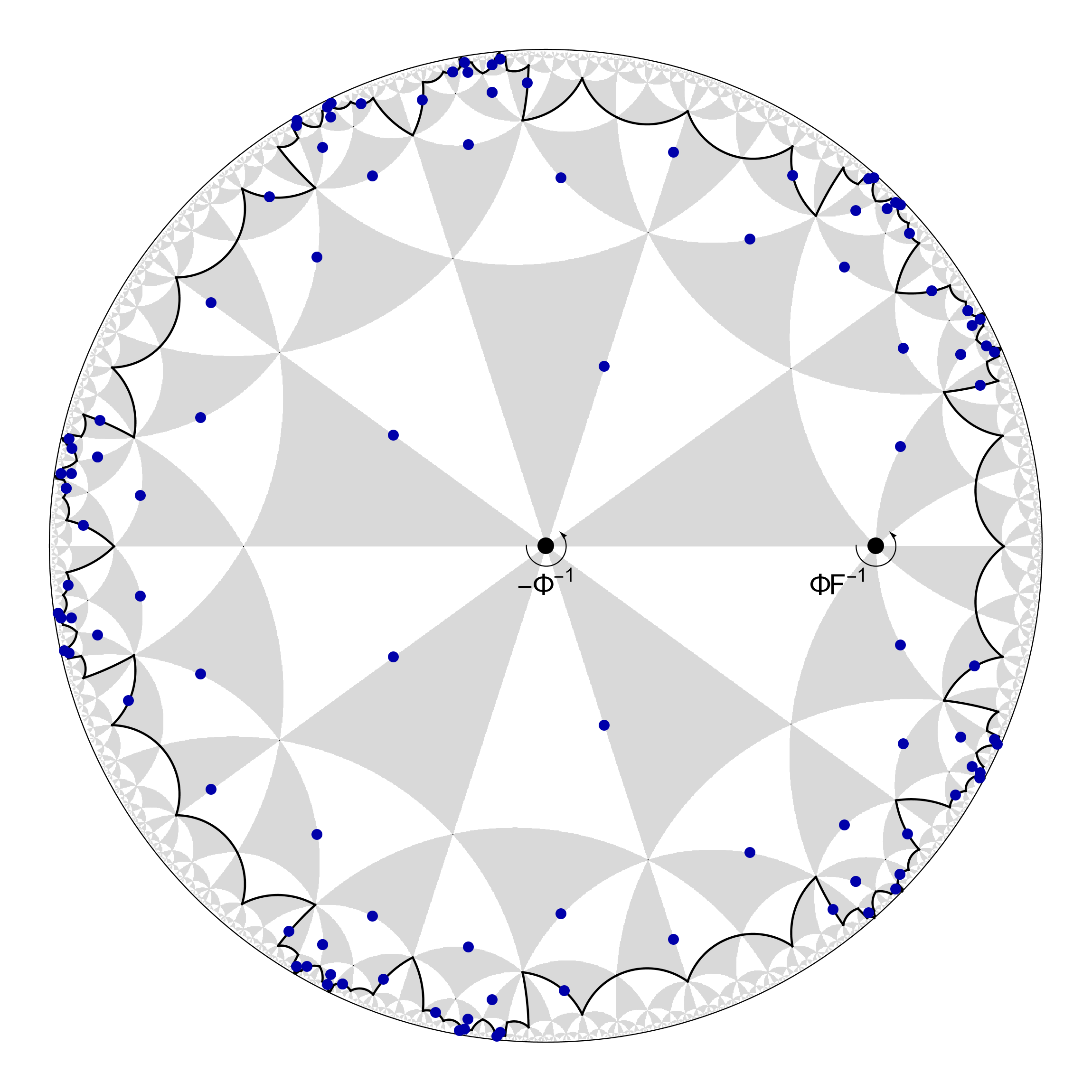}
    \caption{The upper and lower graphs illustrate the codeword configuration of the logical $\ket{\bar{0}}$ and $\ket{\bar{1}}$ states Example \ref{example:4352I} respectively. }
    \label{fig:4352Ihyperbolic}
\end{figure}

\end{appendix}

\end{widetext}

\bibliography{geometry}
\end{document}